\documentclass[10pt]{article}

\usepackage{amsmath}
\usepackage{amssymb}

\usepackage{graphicx}

\usepackage{cite}

\usepackage{color} 
\graphicspath{{./}}

\pdfoutput=1
\usepackage[labelfont=bf,labelsep=period]{caption}

\makeatletter
\renewcommand{\@biblabel}[1]{\quad#1.}
\makeatother

\date{}

\begin{document}

\begin{flushleft}
{\Large
\textbf{Clique of functional hubs orchestrates population bursts in developmentally
regulated neural networks}
}
\vskip 0.5 cm
\textbf{S. Luccioli$^{1,2,\ast}$, E. Ben-Jacob$^{2,3}$, A. Barzilai$^{2,4}$, P.
Bonifazi$^{2,3,\ast,\#}$, A. Torcini$^{1,2,5,\#}$}
\vskip 0.5 cm
\textbf{1} Consiglio Nazionale delle Ricerche, Istituto dei Sistemi Complessi,
via Madonna del Piano, 10, 50019 Sesto Fiorentino, Italy
\\
\textbf{2} Joint Italian-Israeli Laboratory on Integrative Network Neuroscience, 
Tel Aviv University, 69978 Ramat Aviv, Israel
\\
\textbf{3}  Beverly and Sackler Faculty of Exact Sciences School of Physics and Astronomy, 
Tel Aviv University, 69978 Ramat Aviv, Israel
\\
\textbf{4} Department of Neurobiology, George S. Wise Faculty 
of Life Sciences and Sagol School of Neuroscience, Tel Aviv University, Israel
\\
\textbf{5} INFN - Sezione di Firenze and CSDC, via Sansone 1, I-50019 Sesto Fiorentino, Italy
\\
$\ast$ E-mail: stefano.luccioli@fi.isc.cnr.it
\\
$\ast$ E-mail: paol.bonifazi@gmail.com
\\
\textbf{$\#$} Joint Senior Authorship
\end{flushleft}

\section*{Abstract}
It has recently been discovered that single neuron stimulation can impact network dynamics 
in immature and adult neuronal circuits. Here we report a novel mechanism which can explain 
in neuronal circuits, at an early stage of development, the peculiar role 
played by a few specific neurons in promoting/arresting the population activity. For this purpose, 
we consider a standard neuronal network model, with short-term synaptic plasticity, 
whose population activity is characterized by bursting behavior. The addition 
of developmentally inspired constraints and correlations in the distribution of the 
neuronal connectivities and excitabilities leads to the 
emergence of functional hub neurons, whose stimulation/deletion is critical for the network activity. 
Functional hubs form a clique, where a precise sequential activation of the neurons 
is essential to ignite collective events without any need for a specific topological 
architecture. Unsupervised time-lagged firings of supra-threshold cells, in connection with 
coordinated entrainments of near-threshold neurons, are the key ingredients to orchestrate 
population activity.

\section*{Author Summary}

To which extent a single neuron can influence brain circuits/networks dynamics?
Why only a few neurons display such a strong power ?
These open questions are inspired by recent experimental observations in developing
and adult neuronal circuits, as well as by classical debates within the framework of
the single neuron doctrine. In this work we identify and present a mechanism which 
can explain in neuronal circuits, at some early stage of their development,
how and why only a few specific neurons can exhibit such power. For this purpose, we consider 
a standard neuronal network model
whose population activity is characterized by bursting behavior. 
The introduction of a distribution of correlated neuronal excitabilities 
and degrees, inspired by the simultaneous presence of younger and older neurons 
in the network, leads to the emergence of functional hub neurons. 
These critical cells, whenever perturbed, are 
capable of suppressing network synchronization. Notably, we show that their strong influence 
on the population dynamics is not related to their structural properties, but to their
operational and structural integration into a clique. These results highlight how network-–wide 
effects can be induced by single neurons without any need for a specific topological architecture.

\section*{Introduction}

There is increasing experimental evidence that single neuron firing can impact 
brain circuits dynamics \cite{Wolfe2010}. It has been shown that a single pyramidal cell can trigger 
whisker deflection \cite{Brecht2004}, drive sensory perception
\cite{Houweling2007} and modify brain states \cite{Cheng2009}. Similarly, a single 
GABAergic hub cell can affect collective activity within the developing hippocampal 
circuitries \cite{Bonifazi2009}. In vivo cortical studies have shown that 
a single extra action potential (AP) can generate a few dozens extra spikes 
in its postsynaptic targets~\cite{London2010}. Furthermore, a burst of APs, evoked in a pyramidal cell, 
can propagate through the network activating locally a high fraction of Somatostatine GABAergic cells 
(a subset of inhibitory neurons) and a few excitatory cells~\cite{Kwan2012}. 
The capability of single neurons to evoke sparse~\cite{London2010} and network-wide neuronal events 
\cite{Bonifazi2009, Cheng2009, Wolfe2010} in brain circuits can be interpreted within the framework of 
the single neuron doctrine, firstly postulated on sensorial perception by Barlow in 1972~\cite{Barlow1972}. 
According to this doctrine, the spiking of a single neuron in a network
has a high functional relevance being able to code very specifically for high level features of abstraction 
such as concepts. Face selective cells~\cite{Quiroga2005} are a typical example of sparse object representation 
in the brain and of putative ”grandmother” cells~\cite{Perrett1982}.
The sensitivity of neuronal networks to small perturbations, such as those introduced by the firing 
of a single cell, can also find an explanation within the self-organized criticality 
(SOC) framework~\cite{Beggs2003}. In the last decade, SOC has widely been 
proposed as the mechanism underlying power law distributions, 
with characteristic exponents, featuring the size and duration of 
population events. These distributions have been measured in-vivo and in-vitro 
experiments on neuronal networks from invertebrates, rodents, monkeys and humans~\cite{Mazzoni2007,Petermann2009,Hahn2010,Shriki2013}.
The hypothesis underlying the SOC interpretation is that neuronal networks self-organize 
into a critical state where responses, over temporal and spatial scales of any size 
(the so-called ``avalanches''), can be triggered by small perturbations.
Despite the theoretical frameworks above introduced, one of 
the main open question is how and why only specific neurons can affect the global 
network dynamics as observed in ~\cite{Brecht2004, Houweling2007,Cheng2009,Bonifazi2009}.
Two main approaches can be foreseen: a "structural-functional" approach ~\cite{Feldt2011,Bullmore2012,Lee2011,Vasquez2013, Jahnke2013}, 
where the specific topology of the network and the connectivity pattern of the cells 
are responsible for the relevance of the single neuron or a
``dynamical'' approach, where the single neuron becomes relevant due to the nonlinear 
evolution of neuronal excitability and synaptic connectivity in the network~\cite{Wallach2012, Levina2007}.

A recent computational study on the synchronization properties of a specific neural 
circuit~\cite{Gaiteri2011}, has pointed out that the level of burst synchrony is a 
function of both  the network topology and the intrinsic dynamics of peculiar neurons, which 
have a central location in the network graph. This led the authors 
to conclude that in realistic neuronal systems the choice of a specific topology is not sufficient to induce 
an unequivocal  dynamical behavior in network activity. 
To further deepen the comprehension 
of the interplay among cell excitability and synaptic connectivity 
in promoting network burst synchrony, in this paper we study 
the effect of single neurons perturbations 
on the collective dynamics of a network of leaky-integrate-and-fire 
neurons with short-term synaptic-plasticity~\cite{Tsodyks2000Synchrony}.
The relevance of these network models for neuroscience have been demonstrated in many 
contexts ranging from the modelization of working memory~\cite{Mongillo2008}
to the possibility to perform computation by ensemble synchronization~\cite{Loebel2002}.
Although these models have extensively been studied for their capability to generate spontaneous 
population bursting, little is known about the influence of single cell 
perturbations on their global dynamics~\cite{Tsodyks2000Synchrony}.  

In order to analyze the population dynamics in a
neural circuit at the initial stage of its development, when both mature
and young cells are simultaneously present, we consider a random diluted network presenting
developmentally inspired correlations between neuronal excitability and connectivity.
The presence of these correlations can render the network sensitive to single neuron perturbation of a 
few peculiar neurons. The coherent activity of the network can be even arrested by 
removing or stimulating any of these neurons, which are functional hubs
arranged in a {\it clique} regulating the neuronal bursting. 
We show that the level of available synaptic resources influences the
reciprocal firing times of the synaptically connected neurons of the clique.
However, the fundamental mechanism responsible for the burst triggering relies on an 
unsupervised process leading to a precise firing sequence between the neurons which 
are not structurally connected. Furthermore, frequency locking of the same neurons led, counter-intuitively, 
to anti-resonances~\cite{Borkowski2010,Lysyansky2011}, inducing reduced bursting activity
or even complete silence in the circuit. 
Notably, although obtained in a developmentally regulated framework, 
these results can also be extended to a more general context where the effective connectivity and excitability 
of the neurons are dynamically regulated by the different states of brain processing.

\section*{Results}

In this paper we intend to mimic an immature neuronal network {\it frozen} 
at a certain stage of its initial development, similar to the one examined 
in the experimental work on developmental hippocampal circuits~\cite{Bonifazi2009} which 
inspired this work.  At early postnatal stages, the main features characterizing such networks are
the excitatory action of GABAergic transmission (which is instead the most common inhibitory 
source in mature circuits) and the presence of synchronized network events, as 
largely documented in central and peripheral nervous circuits \cite{Allene2008}.
According to that, we consider a network model composed of only excitatory neurons 
and displaying bursting activity. 

In particular, we considered a directed random network made of $N$
leaky-integrate-and-fire (LIF) neurons~\cite{LIFreview1,LIFreview2} interacting via 
excitatory synapses and regulated by short-term-synaptic-plasticity (see Methods for more details), 
similarly to the model introduced by Tsodyks-Uziel-Markram (TUM) \cite{Tsodyks2000Synchrony}.
As previously shown in \cite{Tsodyks2000Synchrony,Loebel2002,Stetter2012}, these networks exhibit a dynamical behavior 
characterized by an alternance of short periods of quasi-synchronous firing
({\it population bursts}, PBs) and long time intervals of asynchronous firing.
Notably, the presence of short-term-synaptic-plasticity is the crucial ingredient to observe 
PBs, even without an inhibitory population~\cite{Tsodyks2000Synchrony,Loebel2002,Stetter2012}.
Therefore, the TUM model with excitatory synapses can be considered as a minimal model to mimic the 
experimentally described stereotypical/characteristic condition of developing neuronal networks~\cite{Ben2002}.

Furthermore, in developing networks, both mature and young neurons are present at the same time, and 
this feature is reflected in the variability of the structural connectivities and of the
intrinsic excitabilities. Experimental observations indicate that
younger cells have a more pronounced excitability~\cite{Ge2005,Doetsch2005},
while mature cells exhibit a higher number of synaptic inputs~\cite{Bonifazi2009, Marissal2012}. 
Thus suggesting that the number of afferent and efferent synaptic connections~\cite{Bonifazi2009,Picardo2011,Marissal2012}
as well as their level of hyperpolarization~\cite{Karayannis2012} are positively correlated with 
the maturation stage of the cells. The gradient of excitability - with younger neurons more excitable than
older ones - could be explained by a gradient in the excitatory action of GABAergic transmission, 
i.e. older neurons receive a less depolarizing action by GABAergic input~\cite{Ben2002}.

The presence at the same time of younger and older neurons can be modeled by considering correlations 
among the in-degree and out-degree of each cell as well as among their intrinsic excitability
and connectivity. In particular, in the attempt to find the network organization 
which is more sensitive to single neuron perturbations, we compare the dynamics of networks where none, 
one or more of the following correlations have been embedded 
(for more details see Methods and Supplementary Information):
 
\begin{itemize}

\item{setup T1:} positive correlation between the in-degree and out-degree of each neuron;

\item{setup T2:} negative correlation between the intrinsic neuronal excitability and the 
total connectivity (in-degree plus out-degree);

\item{setup T3:} positive correlation between the intrinsic neuronal excitability and the 
total connectivity (in-degree plus out-degree). 
 
\end{itemize}

Correlated networks with all possible combinations of the setups T1-T3 have been examined.
However, the paper is mainly devoted to the comparison of the properties of the 
network with correlations of type T1 and T2 (as displayed in Fig. S1) with the 
completely uncorrelated one, which is a directed Erd\"os-R\'enyi graph (see Figs. S3A, S3B).
In order to test the possible influence of hub neurons on the network 
dynamics, also few structural hubs have been added to the network whenever correlations of 
type T1 were embedded (see Methods and  Fig. S1 for more details).

It is important to stress that correlations of type T1 and T2 have a 
justification on the fact that we consider networks at their developmental stage,
as explained above. Furthermore, the correlation of type T2 can also represent a homeostatic regulation
of the neuronal firing to cope with different levels of synaptic inputs~\cite{Turrigiano1998}.
 
For clarity reasons, the paper will mainly deal with a specific realization of a network, made of $N=100$ neurons
and embedding correlations of type T1 and T2. However, we have verified the validity of our findings in other
five realizations of the network with correlations T1 and T2: 
four for $N=100$ (examined in Text S2) and one 
corresponding to $N=200$ (discussed in details in Text S3).

\subsection*{Single neuron stimulation/deletion impacts 
population bursting in developmentally correlated networks}

\begin{figure}[h!]
\begin{center}
\includegraphics*[angle=0,width=16.5cm]{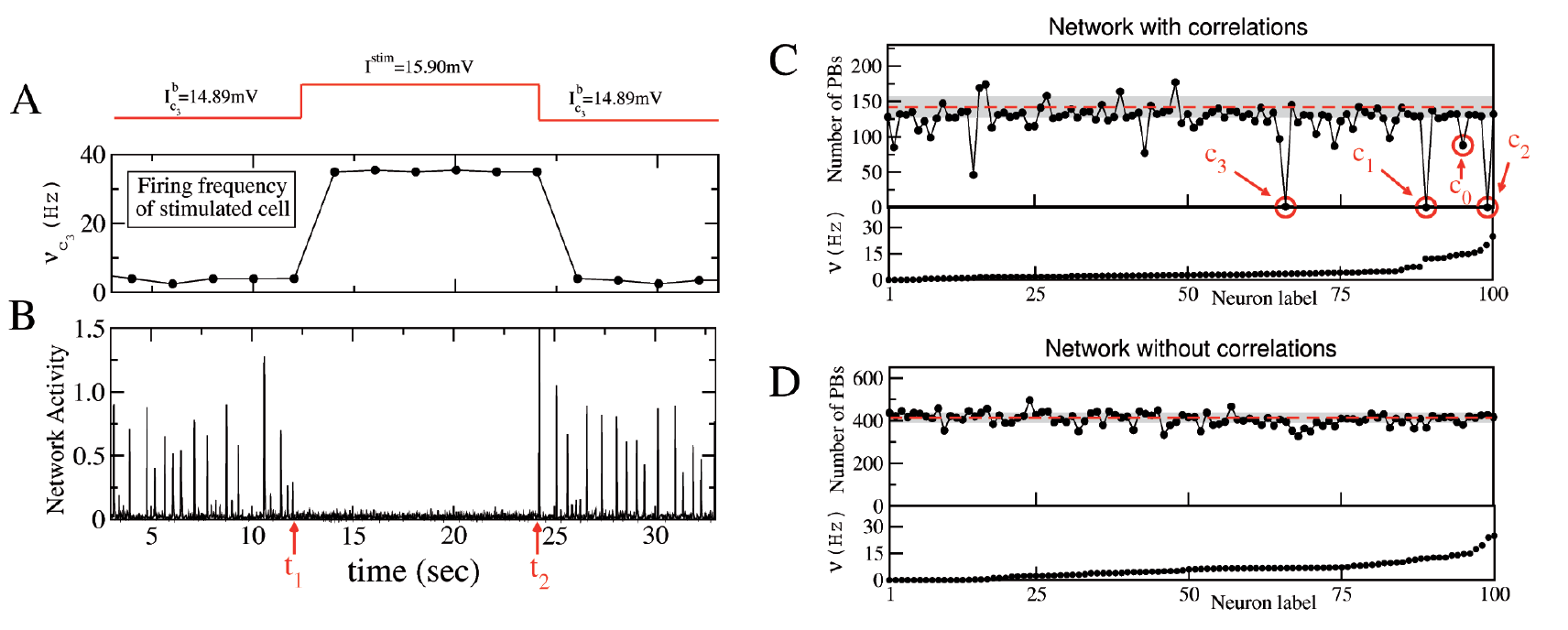}
\end{center}
\caption{
\textbf{Single neuron stimulation (SNS) can stop population bursting activity in 
presence of type T1 plus T2 correlations}. A sketch of a SNS experiment for a network with type 
T1 plus T2 correlations is reported in (A) and (B): the neuron $c_3$ is stimulated with a DC step for a time 
interval $\Delta t=t_{2}-t_{1}$ (as shown by the red line on the top panel). 
Average firing rate of neuron $c_3$ (A) and network activity (B) as 
measured during the experiment. (C) and (D) refer to correlated and uncorrelated networks,
respectively. Upper panels display the number of population bursts, PBs, delivered 
during SNS experiments versus the stimulated neuron, ordered accordingly to their average 
firing rates $\nu$ under control conditions (bottom panels). Each neuron $i$ was stimulated with a DC step 
(switching its excitability from $I_{i}^{b}$ to $I^{{\rm stim}}$) for an interval 
$\Delta t= 84$ s. The critical neurons are signaled by red circles.
The number of PBs, emitted in control conditions within an interval $\Delta t=84$ s, are also displayed:
red dashed lines indicate their averages, while the shaded gray areas correspond to three standard deviations. 
The data refer to $I^{{\rm stim}}=15.90$ mV and $N=100$ neurons.
}
\label{esempioesperimentoBonifazi}
\end{figure}

In the developing hippocampus it has been shown how the stimulation of specific single neurons
can drastically reduce the frequency of the PBs \cite{Bonifazi2009, Feldt2011}. 
These neurons have been identified as {\it hub cells} for their high degree of functional, 
effective, and structural connectivity~\cite{friston1994}. Stimulation consisted of phasic or tonic current 
injection capable of inducing sustained  high firing regime of the 
stimulated neuron over a period of a few minutes. Based on this experimental 
observations, we tested the impact of prolonged {\it single neuron stimulation} (SNS)
on the occurrence of PBs on our network model.
SNS was obtained by adding abruptly a DC current term to the considered neuron. 
For illustrative purpose, we report in Fig.~\ref{esempioesperimentoBonifazi} A-B
the stimulation protocol for a specific neuron capable of suppressing the occurrence of PBs for all the duration of the SNS (in this case limited to $12$ s).
During the current stimulation the neuron fired with a frequency of $\simeq 36$ Hz well above 
the average ($6 \pm 5$ Hz) and the maximal ($24$ Hz) firing rate of the neurons in the network 
under control conditions (see the bottom panel in Fig.~\ref{esempioesperimentoBonifazi} C). 

The stimulation process was completely reversible and after the end of the SNS both the firing rate 
of the cell and the PBs frequency returned to the pre-stimulation control level. 
In order to evaluate the impact of single neuron perturbation on the collective dynamics, we considered 
the variation of PB frequency relative to control conditions (i.e. in absence of any stimulation).
In Figs.~\ref{esempioesperimentoBonifazi} C,D the impact of a single neuron stimulation on the 
PBs frequency is reported for a classical Erd\"os-R\'enyi network (no correlations) and a network with embedded correlations 
T1 plus T2. Please notice that the neurons in Figs.~\ref{esempioesperimentoBonifazi}C-D are ordered according to their 
average firing rate $\nu$ under control conditions, which covered the interval $[0.03;24.80]$ Hz.
The comparison of panels C and D in Fig.~\ref{esempioesperimentoBonifazi} clearly shows that the correlated network 
is much more sensitive to single neuron stimulation. In fact, the SNS was able, for three neurons, to suppress the occurrence of PBs during their stimulation, while for approximately another half dozen of neurons the PBs were halved with respect to control conditions. 
The three most critical neurons $c_1$, $c_2$ and $c_3$ were characterized, before stimulation, by firing frequencies larger than $3.3$ Hz, and they lay among the top $33$\% fastest spiking neurons.  On the contrary, 
in a network where no correlations were present, the SNS had only extremely marginal influence on the 
population activity (see Fig. \ref{esempioesperimentoBonifazi}D), although the firing rate distributions 
in the correlated and uncorrelated network were extremely similar (under control conditions)
as shown in the bottom panels of Fig.~\ref{esempioesperimentoBonifazi} C and D.

In \cite{Tsodyks2000Synchrony} it was shown that the elimination of a pool of neurons from an
uncorrelated network encompassing short-term synaptic plasticity caused a strong reduction 
of the population bursts. In this work we repeated such numerical experiment with single cell resolution, i.e. 
we considered the influence of {\it single neuron deletion}, SND, on the neuronal response of 
the network (results reported in Fig.~\ref{TUMexperiment}). 
For the network with correlations $T1$ plus $T2$, in four cases the SND led to the 
complete disappearance of PBs within the examined time interval, while for five other neurons 
their individual removal led to a decrease of the order of $\simeq 30-40 \%$ in the frequency of occurrence of 
the PBs (Fig.~\ref{TUMexperiment} A). Three among these four critical neurons (namely, $c_1$, $c_2$ and $c_3$)
were also responsible for silencing the network during the SNS experiment performed
with a stimulation current $I^{stim}=15.90$ mV, as shown in Fig. \ref{esempioesperimentoBonifazi}C 
and Fig. \ref{TUMexperiment} A. The same SNS experiment on the fourth critical neuron,
labeled $c_0$, reduced the PB frequency of about 40\% with respect to control conditions
(Fig. \ref{esempioesperimentoBonifazi}C). However, as we will show in the following, a SNS
experiment performed on $c_0$ with a different injected current can also lead to a complete silence 
in the network. Notably, in analogy with the SNS, also this additional critical neuron $c_{0}$ impacting the 
PB occurrence under SND lays among the top 33\% fastest spiking neurons.  
Differently from SNS, the removal of neurons with lower frequencies had almost no impact 
on the network dynamics. For uncorrelated networks the effect of SND was almost negligible, inducing a maximal 
variation in the bursting activity of the order of 10-15 \% with respect to the activity under control conditions (see Fig. S3C).

\subsubsection*{Other Correlation Setups}

 We have also analyzed the response to SNS and SND experiments in networks embedding
all the possible combinations of the correlations setups T1-T3. In particular, we considered
networks with positive correlation between structural in-degree ($K^I$) and 
out-degree ($K^O$) (setup T1 shown in Figs. S4A, S4B), with negative correlation
between excitability $I^b$ and total connectivity $K^T$ (setup T2 shown in Figs. S5A, S5B), with only positive correlation between $I^b$ and $K^{T}$ 
(setup T3 shown in Figs. S6A, S6B) and finally combining positive correlations between $K^I$ and $K^O$ and $I^b$ and $K^{T}$ 
(setups T1 plus T3 shown in Figs. S7A, S7B). 

As we are looking for strong impacts on the network dynamics, we identified
the network as ``sensitive'' to SND (or SNS) whenever the PBs frequency was altered
more than 90\%  with respect to the corresponding PB activity in control conditions.
Therefore, we considered as significative only the modifications of the activity 
which were well beyond the statistical fluctuations in the population bursting, shown by the shaded gray area 
in panels C and D in Figs. S4, S5, S6, and S7.
 
In all the examined cases, despite the fact that the firing frequencies distributions 
were quite similar to the ones measured in the correlated network embedding setups T1 and T2, 
we did not observe significant modifications of the bursting activity by performing SNS and SND experiments 
on any neuron of the network (see panels C and D in Figs. S4,S5,S6,S7). The situation 
where SNS and SND had a larger effect on the network activity was for the correlations of type T2. 
In that specific case we observed that SND on 2 neurons (lying among the top 33\% fastest spiking neurons, 
namely $\nu > 5$ Hz) halved the bursting frequency of the network, and SNS on one of these 2 neurons had 
a similar effect. In all the other cases the PB activity was never perturbed more than 20-30 \%.
Only the simultaneous presence of type T1 and type T2
correlations noticeably enhanced the sensitivity to SNS and SND, 
leading to the possibility to silence the network.

\subsection*{Structural and functional properties of the network}

In order to gain some insight into the mechanisms underlying the reported 
response of the network, with correlations of type $T1$ plus $T2$, to SNS and SND experiments, 
we analyzed the structural and functional connectivity  
of the network in relation to the intrinsic excitability of the neurons. 
Functional connectivity (FC) analysis~\cite{Bullmore2009} was aimed at revealing 
time-lagged firing correlations between neuronal pairs, similarly to what described 
in~\cite{Bonifazi2009} for the developing hippocampus.
In particular, for every possible pairs of neurons $(i,j)$ we cross-correlated their spike 
time series, with the exclusion of the spikes occurring 
within bursts, for which only the timestamp of the first spike was kept (see Methods). 
A functional connection directed from $i$ to $j$ was established whenever the activation of $i$ reliably preceded 
the activation of $j$ and viceversa (see Methods). For each cell $i$, we calculated the functional out-degree 
(in-degree) $D^O_i$ ($D^I_i$), i.e. the number of cells which were reliably activated after (before) 
its firing.

\begin{figure}[h!]
\begin{center}
\includegraphics*[angle=0,width=16.3cm]{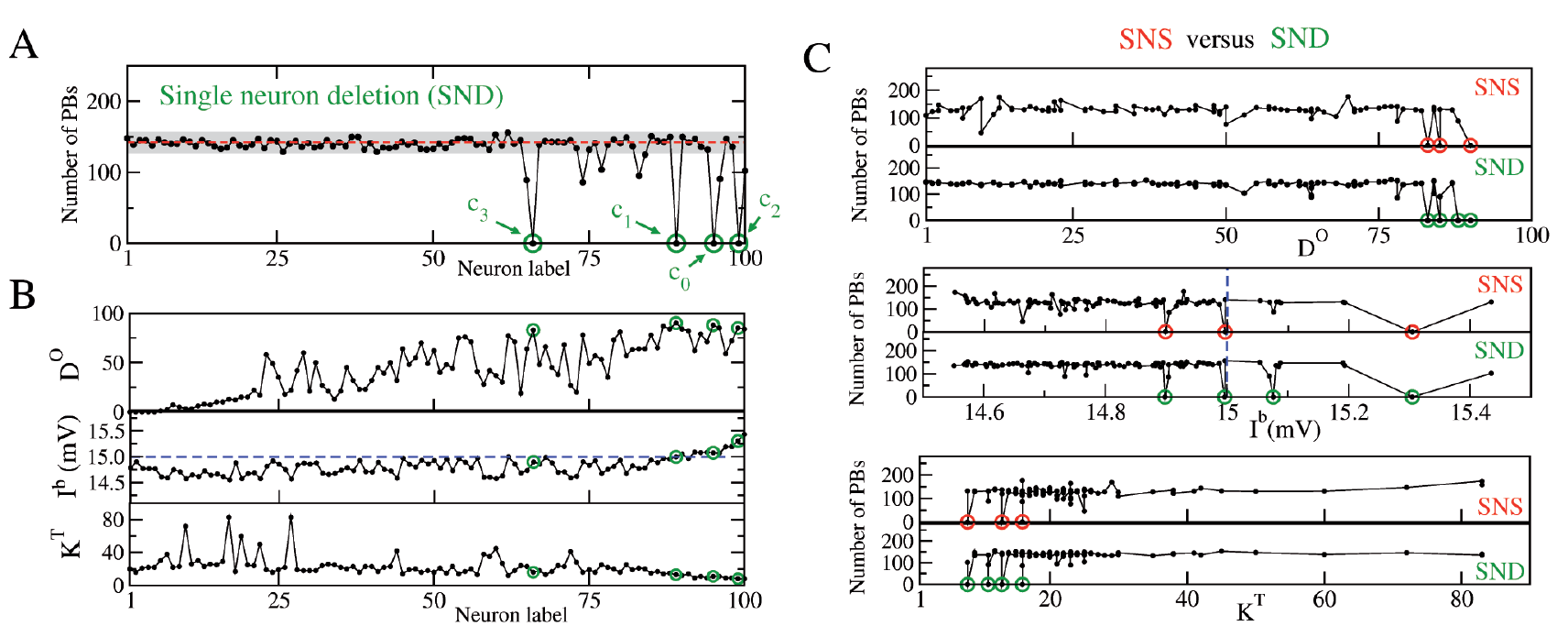}
\end{center}
\caption{{\bf Comparison between single neuron stimulation (SNS) and deletion (SND) 
in a network with correlations of type $T1$ plus $T2$.}
(A) Number of PBs emitted during SND experiments versus the label of the removed neuron. 
(B) Functional and structural properties of the network, as measured in
control conditions, i.e. in absence of any stimulation/manipulation of the neurons.
From top to bottom: functional out-degree $D^0$, 
intrinsic excitability $I^b$, and total structural connectivity $K^{T}$.
The red dashed line and the gray shaded area in (A) as well as the neuron labels
are as in Fig.~\ref{esempioesperimentoBonifazi} C, the blue dashed line denotes
$V_{th} = 15$ mV. (C) Comparison between SNS and SND: 
the number of PBs occurring during SNS (resp. SND) is reported as a function 
of $D^0$, $I^b$ and $K^{T}$. In all panels the green (red) circles mark the critical 
neurons, which under SND (SNS) can silence the bursting activity of the network.
The bursting activity is recorded over an interval $\Delta t=84$ s.
}
\label{TUMexperiment}
\end{figure}

As shown in the top panel of Fig. \ref{TUMexperiment}B, the four critical neurons, $c_0$--$c_3$,
identified during the SNS and SND experiments, have very high functional out-degree, namely
$83 \le D^O_i \le 90$. In particular, three of them ($c_0$, $c_1$ and $c_2$) are 
ranked among the first four neurons with the highest functional out-degree.
Therefore the critical neurons are reliably preceding the activation of most 
of the other neurons in the network. In addition,  neurons $c_0$ and $c_2$ were 
supra-threshold ($I^b > V_{th}$, see Methods) and therefore firing tonically 
even if isolated from the network, while neuron $c_1$ was at threshold and $c_3$ 
below it (as shown in the central panel of Fig.~\ref{TUMexperiment}B).

In contrast to their high functional out-degree, critical neurons were characterized by a low structural 
degree $K^T$ (total number of afferent and efferent connections), namely
$K^T < 16$ with respect to an average value $23 \pm 13$, as shown in the bottom panel of Fig.~\ref{TUMexperiment}B.
This result was a direct consequence of the anti-correlation imposed between total degree and excitability
and this represented a crucial aspect for the emergence of the critical neurons.
 
In Fig.~\ref{TUMexperiment} C we report the results of SNS (SND) experiments
as a function of $D^0$, $I^b$ and $K^T$ of the stimulated (removed) neurons.
The experiments on the neurons with high $K^{T}$ 
(the structural hubs, shown in Fig. S1 A and S1 B) influenced marginally the network bursting,
apart for the single neuron stimulation of the two principal hubs which led to a moderate increase of the 
activity (see the bottom panels in Fig.~\ref{TUMexperiment} C). However SND on the same neurons had
no significant effect.  On the other hand, neurons with high functional out-degree  $D^0$ 
(functional hubs) were quite relevant to sustain the collective dynamics.
The removal of neurons with low $D^0$ (including the structural hubs) 
seemed almost not affecting the bursting properties of the
network. Altogether, apart the stressed differences, the SNS and SND experiments appeared to 
give quite similar results.

The generality of these findings have been tested by performing SNS/SND 
experiments on other five different realizations of the network with embedded correlations of type T1 and T2,
in all cases a small subset of neurons resulted to be critical in the sense discussed above
(for more details see Text S2 and S3).

\subsection*{Network response during SNS: dependence on the injected current}

\begin{figure}[h!]
\begin{center}
\includegraphics*[angle=0,width=16.3cm]{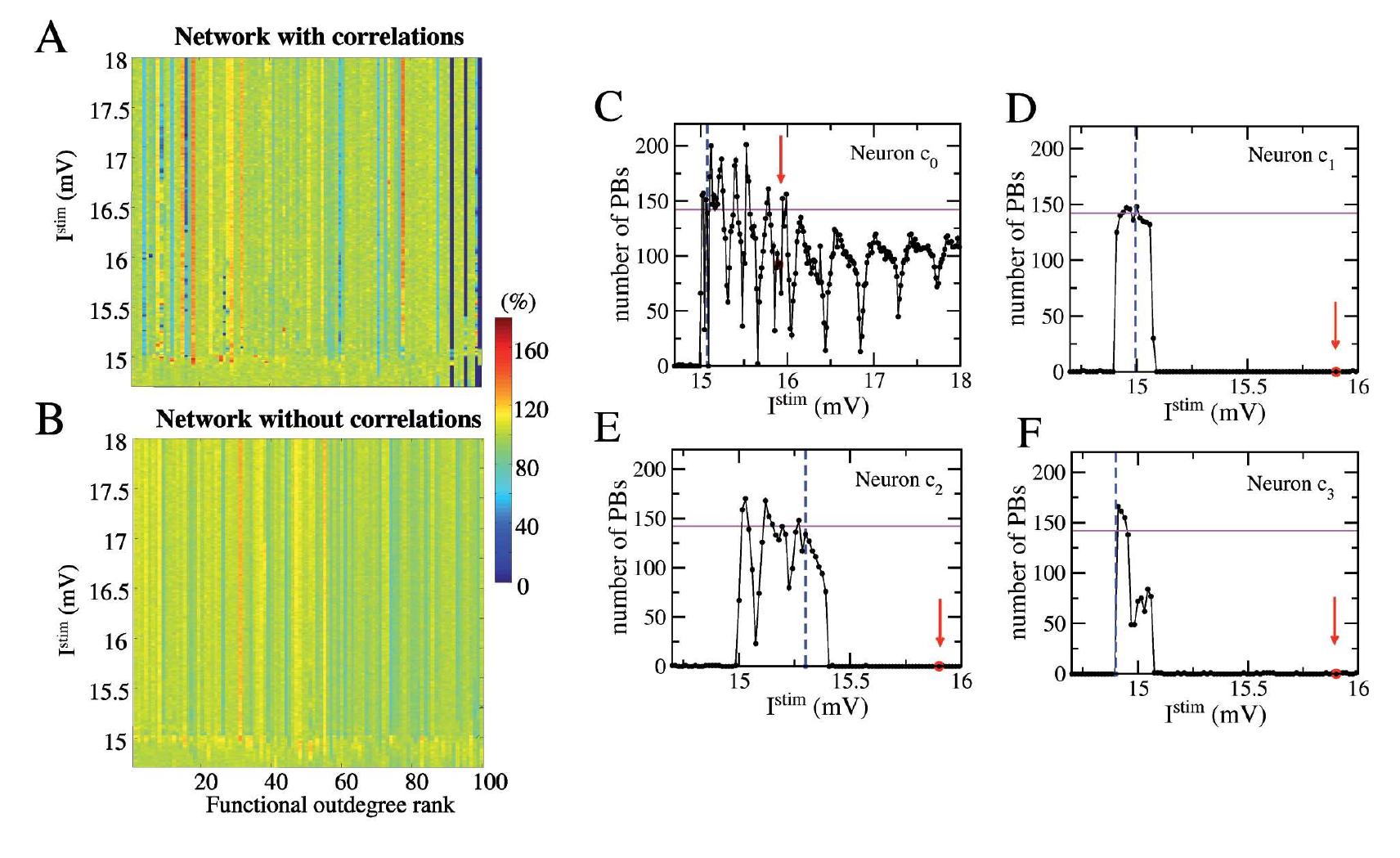}
\end{center}
\caption{{\bf Impact of single neuron stimulation on the population activity: dependence on the injected current.} 
Color coded rates of emission of PBs during SNS experiment performed 
on each single neuron for a range of injected DC currents $I^{stim}$ (y-axis) 
in networks with correlations of type T1 plus T2 (A) and without any correlations (B). 
The neurons are ordered according to their functional out-degree rank (x-axis) and
the PB rates during SNS are normalized to the PB rate in control conditions.
(C-F) Number of PBs emitted during SNS of 
the critical neurons $c_0$,$c_1$,$c_2$,  $c_3$ versus the stimulation current $I^{stim}$ .
The red arrows indicate $I^{stim}$ employed for the SNS experiments
in Fig.~\ref{esempioesperimentoBonifazi} C. The blue vertical dashed lines mark the value 
of the intrinsic excitability and the horizontal magenta solid line the 
bursting activity of the network, both measured at rest. 
The number of PBs are measured over a time interval $\Delta t=$ 84 s.
} 
\label{impatto_di_a_suicritici}
\end{figure}

In order to further clarify the impact of varying the intrinsic excitability of single neurons on the
network bursting activity, we have performed extensive analysis of the network 
response under SNS experiments for 
a wide range of stimulation currents, namely  $I^{stim} \in [14.7 ; 18.0]$ mV. 
In panels A and B of Fig.~\ref{impatto_di_a_suicritici} it is summarized the impact on the bursting activity 
of the SNS for networks with type T1 and T2 correlations and without any correlations. SNS had really a minimal 
effect on the uncorrelated network: in this case the number of emitted PBs varied only up to a 20 \% with 
respect to control conditions. On the contrary, for the correlated network, SNS was able to silence the network
over a wide range of currents when $c_1$, $c_2$ and $c_3$ were stimulated. For the other neurons, 
SNS with high stimulation currents could also have the effect of promoting an increase of PBs up to 
130-140 \% with respect to control conditions.
In particular, an increase in the PB activity has been observed consistently for two structural hubs,
whenever they are brought above the firing threshold, as shown in Fig. S1 C, and for other two neurons
directly connected to these hubs. This behaviour is expected for an excitatory network without correlations, 
where the neurons with higher out-degree have usually the highest impact on the network~\cite{Vasquez2013}.
However, the removal of each of these four neurons from the network did not influence the PB activity,
furthermore they were passively recruited during bursting events.These results had an explanation
in the fact that in control conditions the structural hubs were well below threshold, due to the anti-correlation between total degree 
and excitability, while by increasing the stimulation on these hubs we violated such 
constraint.

As shown in panels D, E, F of Fig.~\ref{impatto_di_a_suicritici}, for neurons  $c_1$, $c_2$, and $c_3$  
the bursting activity survived only in narrow stimulation windows located around, or just above the 
firing threshold value. A current variation $\Delta I^b \simeq 0.1-0.3$ mV was, for these three neurons, sufficient 
to silence the network. The stimulation of the neuron $c_0$ , the one critical for SND but not for SNS 
(when $I^{{\rm stim}}=15.90$ mV), revealed the 
existence of very narrow {\it anti-resonance} windows (i.e. minima in the number of emitted PBs), as shown in Fig.~\ref{impatto_di_a_suicritici} C.
For very specific intrinsic excitability this neuron could effectively silence the network, but 
for generic excitation its influence on PBs activity was limited. The anti-resonances occurred (for $I^b > 15.30$ mV) 
at almost regular intervals: initially of width $\simeq 0.2$ mV and, at larger intrinsic excitability, of width $\simeq 0.4$ mV.
This point will be further discussed and clarified in Sect. {\it  Time Orchestration}.

\subsection*{The functional clique}

\begin{figure}[h!]
\begin{center}
\includegraphics*[angle=0,width=7.9cm]{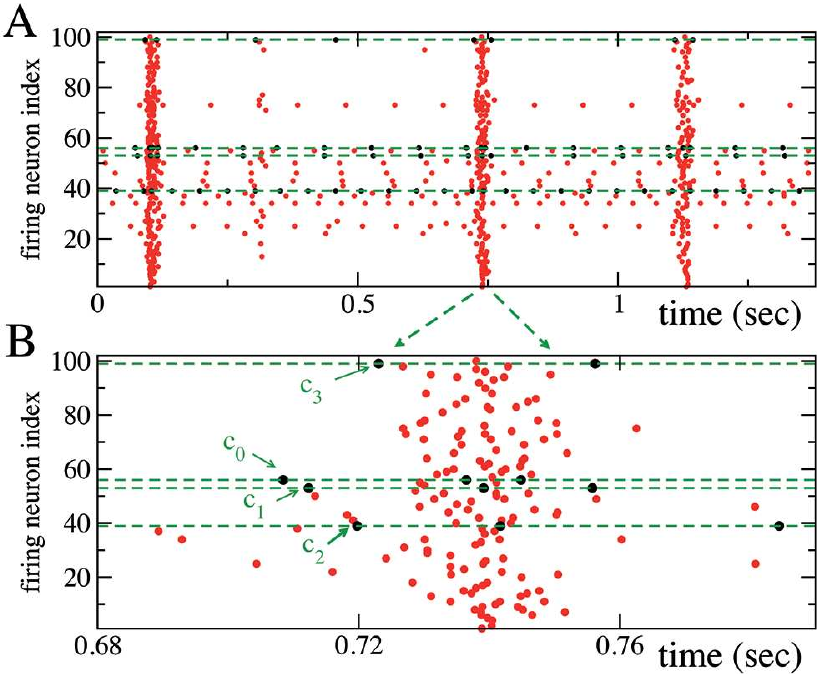}
\end{center}
\caption{{\bf The critical neurons precede the population bursts in
a network with correlations of type $T1$ plus $T2$.}
(A) Raster plot of the network activity: every dot denotes a firing event.
The dashed green lines and black dots refer to the four critical neurons.
(B) Enlargement of a representative population burst: PBs are anticipated by
the ordered firing sequence
$c_0 \to c_1 \to c_2 \to c_3$. For clarity reasons, in the raster plots, at
variance with all the other
figures, the neuronal labels are not ordered accordingly to their firing
rates.
}
\label{burst_synapse}
\end{figure}

The results reported above suggest that the four critical neurons $c_0$,$c_1$,$c_2$,$c_3$, identified in 
the network with correlations T1 plus T2 should have a key role in the onset of the collective bursting. 
Therefore, we focused our analysis on the PB build up, i.e. we examined the events occurring in a
time window of $25$ ms preceding the peak of synchronous activation (for more details see Methods).
In particular, we quantified how many times each single neuron participated in the build up of a PB. 
As we have verified, for the correlated network all the bursts were preceded 
by the firing of the four critical neurons, while in absence of correlations
there was no neuron capable of reliably preceding every burst activation.
The cross correlations between the timing of the first spike emitted by each critical 
neuron during the PB build up (see Methods) are shown in Fig.~\ref{clique} A (blue histograms).
This analysis revealed a precise temporal sequence in the neuronal activation, respectively 
$c_0 \to c_1 \to c_2 \to c_3$, as shown also for a few representative bursts in Fig. \ref{burst_synapse} A,B 
(therefore the labeling assigned to these neurons).
Interestingly, the same neurons did not show this precise temporal activation out of the PBs, as 
revealed by the red histograms in Fig.~\ref{clique} A (see also Methods). 
Furthermore, the time sequence of the firing events of the critical 
neurons during the build up of the PB was quite well determined: $c_0$ anticipated the firing of $c_1$
of $3.94 \pm 0.5$ ms, $c_1$ anticipated $c_2$ of $9.6 \pm 3.3$ ms and $c_2$ anticipated $c_3$
of $3.3 \pm 1.0$ ms. During the inter-burst periods we observed clear time lagged correlations 
only for the pair $c_0 \to c_1$, presenting a direct synaptic connection, and in a weaker manner 
also for the pair $c_2 \to c_3$.
On the basis of the reported data, we can safely affirm that the critical neurons 
form a {\it functional clique} responsible for the onset of the PBs.

\subsection*{The role of plasticity}

As clarified in \cite{Tsodyks2000Synchrony}, the bursting activity was
due to the short-term-synaptic depression. In particular PB emission could be related to 
the evolution of the fraction of synaptic resources in the recovered state, characterized
by the variable $X_i^{IN}$ ($X_i^{OUT}$), averaged over the afferent (efferent) synapses of
each neuron $i$ (see Methods). The authors in \cite{Tsodyks2000Synchrony} have shown
that the fraction of synaptic resources, averaged over all the excitatory synapses,
had a deep minimum in correspondence of the burst event and then slowly recovered 
its stationary value over a time scale dictated by the average recovery time ${\bar T^R}$. 
This means that the average effective strength of the excitatory connections (measured
by $X_i^{IN}$ and $X_i^{OUT}$) was strongly depressed after a burst, and this inhibited the
prosecution of the bursting activity, which could restart only when the strengths
of the synapses would return to their stationary values.
 
\begin{figure}[h!]
\begin{center}
\includegraphics*[angle=0,width=16.3cm]{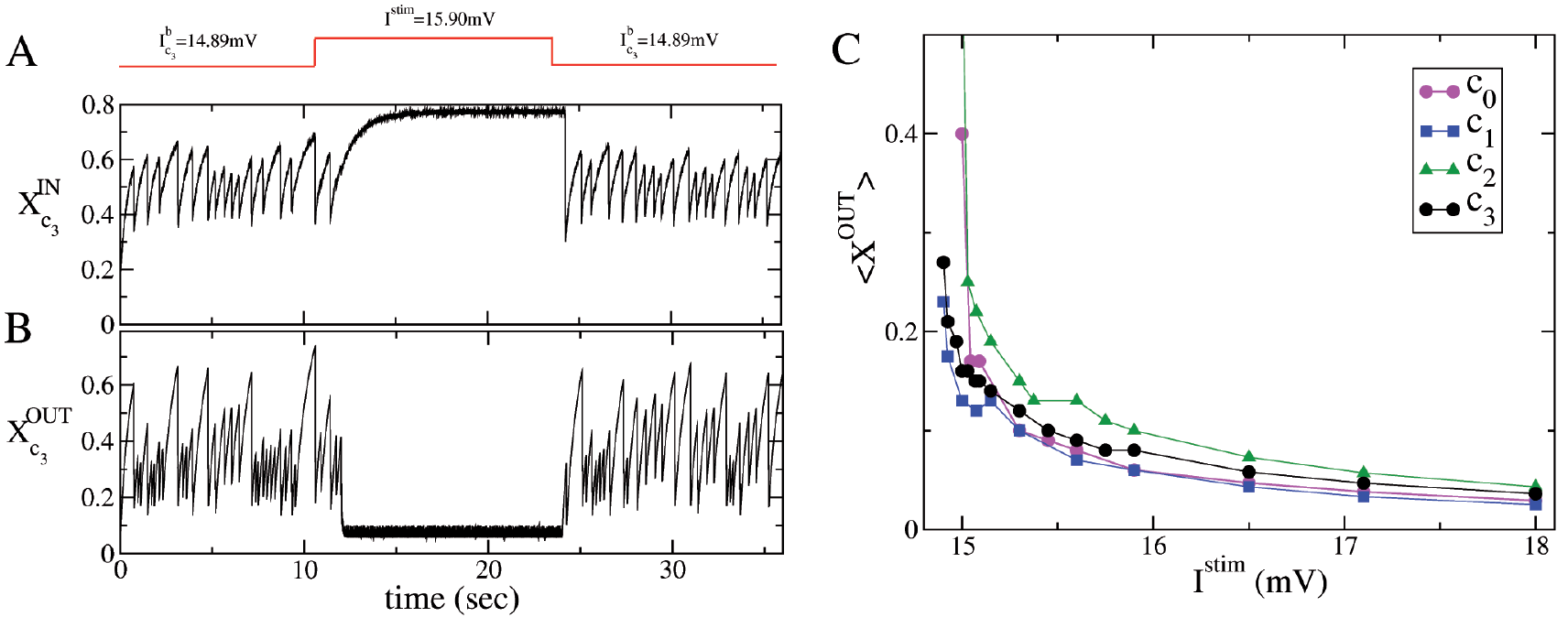} 
\end{center}
\caption{{\bf Effective synaptic strength during single neuron current stimulation.}
Average synaptic strength of the afferent (A), and efferent
(B) connections of the  critical neuron $c_3$ during SNS with $I^{stim} = 15.90$ mV
(these data corresponds to the experiment reported in Fig.~\ref{esempioesperimentoBonifazi}). 
The output (input) effective synaptic strength 
is measured in terms of the average value of the fraction $X^{OUT}_{c_3}$ ($X^{IN}_{c_3}$) of the synaptic 
transmitters  in the recovered state associated to the efferent (afferent) synapses (see Methods).
(C) Time averaged synaptic strengths $\langle X^{OUT} \rangle$ as measured during SNS experiments performed on
each of the four critical neurons for various stimulation currents $I^{stim}$. The 
legend clarifies to which neuron corresponds the average synaptic strengths displayed in the
figure, the averages have been performed over 84 s.
}
\label{effective_synapse}
\end{figure}

In this Section, we want to address the question whether the variation of the effective strength of the 
synapses could be also responsible for the silencing of the network (with correlations of type T1 plus T2)
during SNS experiments. So far we have clarified that the removal of any of the four neurons in the functional clique
blocked the bursting activity, however it is not clear why a small stimulation of $c_1,c_2,c_3$
was capable also of blocking the PBs. 
As reported in Fig.~\ref{effective_synapse} A, B the stimulation
of neuron $c_3$ with a large current $I^{stim} = 15.9$ mV (as in the experiment reported in
Fig.~\ref{esempioesperimentoBonifazi} A,B) reduced noticeably $X^{OUT}_{c_3}$, due to the high
firing activity of the stimulated neuron. Analogous results have been found for all the other
three critical neurons. For neurons $c_1$, $c_2$ and $c_3$ this stimulation blocked the
bursting activity of the network, thus inducing an almost complete 
recovery of the available resources 
of the afferent synapses, measured by $X^{IN}$ (as shown in 
Fig.~\ref{esempioesperimentoBonifazi} A for $c_3$). 
These results could  suggest that SNS and SND experiments are indeed equivalent, 
since if the efferent synapses are extremely  depressed, this could correspond 
somehow to remove the neuron from the network. However, 
SNS of neuron $c_0$ did not lead generically to the suppression of the bursting activity 
even if its efferent synapses were similarly depressed (as shown in Fig.~\ref{impatto_di_a_suicritici} C).
Furthermore, the synaptic depression could not explain the anti-resonances in the 
bursting activity observed for SNS
of $c_0$ and $c_2$ with different $I^{stim}$. Since the time averaged synaptic 
strength, $\langle X^{OUT}\rangle$, exhibited only a smooth decrease as a function of $I^{stim}$ for all the four critical neurons 
(as well as for any generic neuron in the network), as shown 
in Fig.~\ref{effective_synapse} C. 
 
\subsection*{Time orchestration}

\begin{figure}[h!]
\begin{center}
\includegraphics*[angle=0,width=12.0cm]{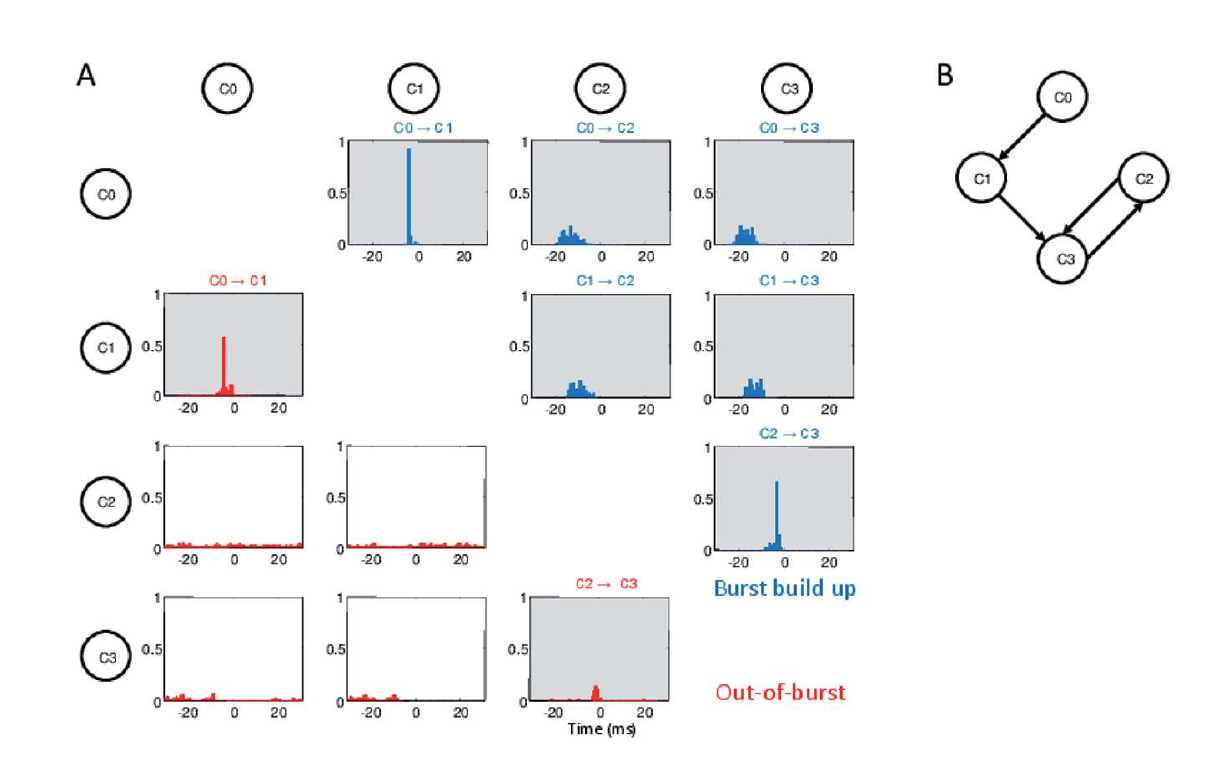}
\end{center}
\caption{{\bf The functional clique}.
(A) Distributions of the firing time delays $\tau_{max}$ between two critical 
neurons. $\tau_{max}$ has been measured as the position of the maximum of the cross correlation
between the time series of the two considered neurons. The panels refer to all the possible
pair combinations of the critical neurons, furthermore blue (red) histograms 
refer to the analysis performed during the population burst build up 
(during periods out of the bursting activity). For more details see the subsection Functional Connectivity 
in Methods. The order of activation of each pair is reported on the
top of the corresponding panel, whenever the cross-correlation has a significant maximum
at some finite time $\tau_{max}$. Note that during the PB onset, neurons activate reliably in the following order 
$c_0 \rightarrow c_1 \rightarrow c_2 \rightarrow c_3$.
During the out-of-burst activity, clear time-lagged activations are present only among the 
pairs $c_0$-$c_1$ and $c_2$-$c_3$. (B) Structural connections among the four critical neurons: 
the black arrows denote the directed connections. The data here reported, as well in all the following 
figures, refer to a network with correlations of type T1 plus T2.
}
\label{clique}
\end{figure}

As already mentioned, the roles of the four neurons in the functional clique 
of the network with type T1 and T2 correlations were quite well
established, and just a precise firing time sequence could induce the population
avalanche. To better understand the role of each critical neuron, 
it is necessary to point out that, under control conditions, 
the neurons $c_0$ and $c_2$ could fire even if isolated (since $I_{c_0}^b = 15.07$ mV
and $I_{c_2}^b = 15.30 $ mV were larger than $V_{th}$), $c_1$ was at threshold 
($I^b_{c_1} = 14.99$ mV) and $c_3$ was the only neuron below threshold 
($I^b_{c_3} = 14.89$ mV). This clearly explains, 
given the existing synaptic connection from $c_0$ to $c_1$ (see Fig.~\ref{clique} B), 
the reason why $c_0$ entrained $c_1$, both during the burst build up 
as well as during the inter-burst periods (see Fig.~\ref{clique}). 
Furthermore, from the results of the SNS experiments performed on $c_1$ and $c_3$ 
(Panels D and F in Fig.~\ref{impatto_di_a_suicritici}) one can observe that the network 
activity arrested whenever $I^{stim} > I_{c_0}^b = 15.07$ mV for both these neurons 
(for comparison, note that the range of $I^{stim}$ reported in panel C is different from panel D,E and F).
Therefore, whenever these two neurons fired faster than the clique leader $c_0$, the burst activity, 
which should be triggered by a well determined sequence of events, would be terminated. Thus we can
conclude that $c_1$ and $c_3$ could only be the followers of the dynamics dictated by the two supra-threshold 
neurons, and in particular by the leader $c_0$.

\begin{figure}[h!]
\begin{center}
\includegraphics*[angle=0,width=16.3cm]{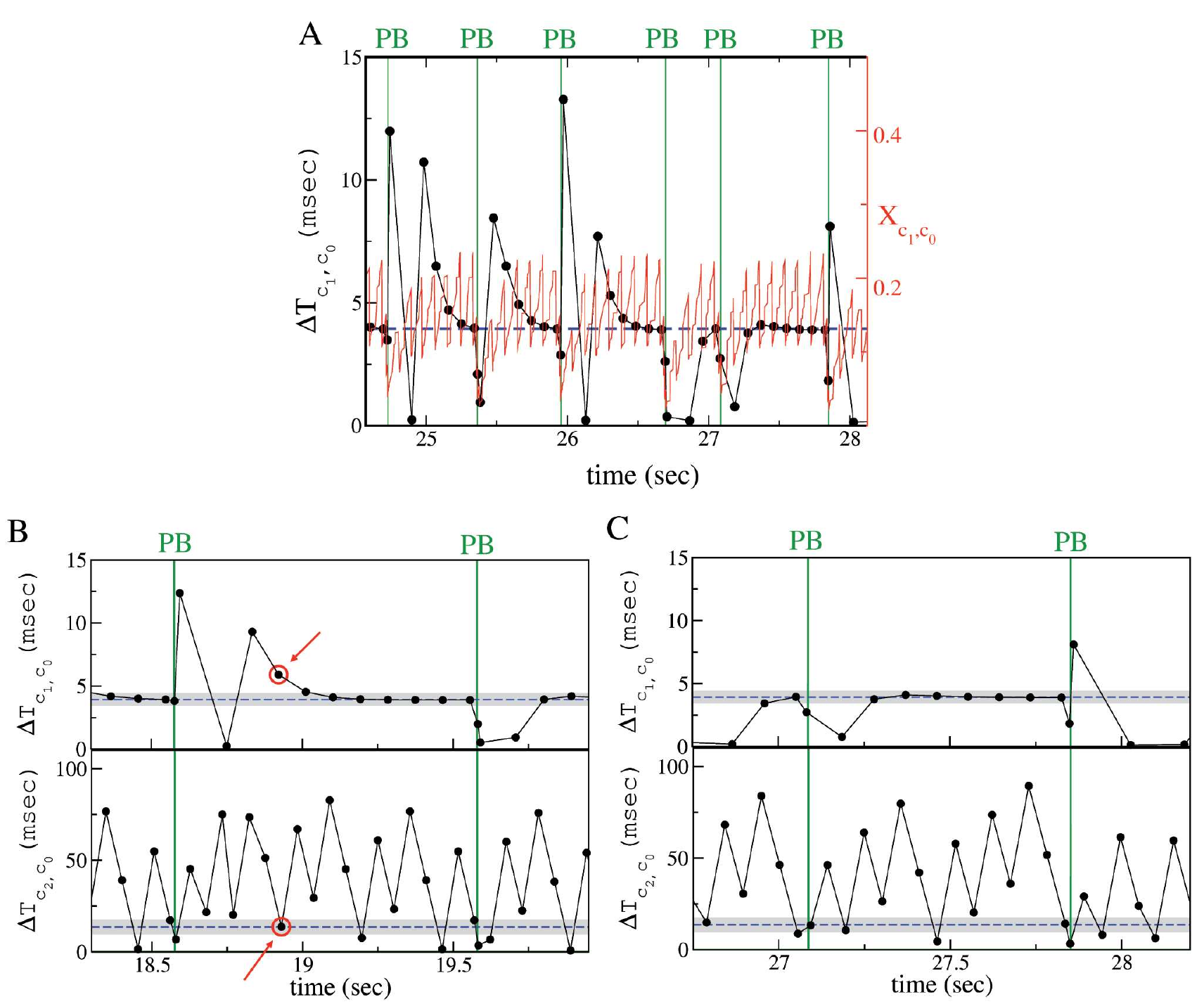}
\end{center}
\caption{ {\bf (A) Synapse strength and firing time delay between the neurons $c_0$ and $c_1$.}
Time evolution of the effective synaptic strength $X_{c_1,c_0}$  (red solid line and right {\it y}-axis) 
and of the firing time delay $\Delta T_{c_1,c_0}$ (black line with dots and left {\it y}-axis).
{\bf (B),(C) Failures and successes in population burst ignition.}
Spike time delay $\Delta T_{c_1,c_0}$ (top panel) and $\Delta T_{c_2,c_0}$ (bottom panel)
of neuron $c_1$ and $c_2$, respectively, referred to the last firing time of $c_0$. Panels (B) and (C) 
clearly show that PBs (denoted by green vertical lines) can occur only when
the neuron $c_1$ and $c_2$ fire within precise time windows after the firing of neuron $c_0$.
In (B) a clear failure is indicated by red circles, in this case $c_2$ fired at
the right time, but $c_1$ was too slow; in (C) neuron $c_1$ fires at the right moment
several times (black dots are within the gray shaded area in the top panel), but the avalanche is not 
initiated until $c_2$ does not emit a spike within 
a precise time interval after the firing of $c_0$. In all the figures, the data refer to control conditions.
The blue horizontal dashed lines refer to the average value of $\Delta T_{c_2,c_0}$ or 
$\Delta T_{c_1,c_0}$ at the PB onset, while the shaded gray areas indicate the corresponding 
standard deviations.
}
\label{PBrec}
\end{figure}

As clearly shown in Fig.~\ref{PBrec} A, 
exactly before a burst event (i.e. in the PB build up phase) neuron $c_1$ 
fired with a precise time lag after neuron $c_0$ (blue dashed line in the figure). However, 
the time lag $\Delta T_{c_1,c_0}$ between the firing 
of $c_0$ and $c_1$ needed some time after each bursting event to adjust 
to its pre-burst value. This could be interpreted also as an effective refractory 
period needed to the pair $c_0$-$c_1$ to recover the proper entrainment favorable
to the burst discharge. As shown in Fig.~\ref{PBrec} A, the time evolution of 
the variable $X_{c_1,c_0}$, which measured the effective strength of the synapse 
connecting $c_0$ to $c_1$, is directly connected to the duration of the time interval $\Delta T_{c_1,c_0}$
(or analogously to the effective refractory time of the entrainment $c_0$-$c_1$).
After a burst, $X_{c_1,c_0}$ was noticeably depressed (reaching almost zero) and it slowly recovered 
its asymptotic value over a time scale dictated by ${T^R}_{c_1,c_0}$. Indeed 
$X_{c_1,c_0}$ was strongly oscillating due to the firings of $c_0$, however the recovery of the
pre-burst condition can be assessed by considering its extreme values (minima and maxima)
both slowly increasing after the burst. The recover of the effective 
synaptic strength was associated to the adjustment of $\Delta T_{c_1,c_0}$ to the 
value taken during the build up of a PB. From Fig.~\ref{PBrec} A, it is also evident that the 
fulfillment of this condition was not sufficient to induce another PB, since the PB could occur 
even a long time after the favorable pre-burst value was reached by $\Delta T_{c_1,c_0}$.

Similar behaviors had been observed also for the
synapse connecting $c_2$ to $c_3$, although the firing of neuron $c_2$ alone was not sufficient 
to bring $c_3$ above threshold and therefore to initiate the PB. 
Indeed, the activation of $c_3$, whose firing was fundamental to trigger the avalanche, 
was more complex. 
From a structural point of view, the neuron $c_3$ received inputs directly from $c_1$ and $c_2$, 
while there were no synaptic connections between $c_1$ and $c_2$ (see Fig.~\ref{clique} B). 
The entrained firing of the pair $c_0-c_1$ followed by the firing of $c_2$, within a precise time window, 
was required to induce $c_3$ to emit a spike and therefore a PB. This can be clearly appreciated 
from Fig.~\ref{PBrec} B and C. In particular, in Fig.~\ref{PBrec} B is reported a situation where
$c_2$ fired at the right time after $c_0$, but $c_1$ has fired too late to start an avalanche in the
network (as previously explained the firing of $c_1$ was not yet entrained to that of $c_0$). 
Much more common is the situation reported in Fig.~\ref{PBrec} C, where $c_1$ fired essentially
always at the same time after $c_0$, but instead the time delay $\Delta T_{c_2,c_0}$ 
in the firing of $c_2$ was extremely variable ranging from an almost coincidence with $c_0$ to a delay of 
100 ms. The PB could occur only when $c_2$ fired in a precise time window following the activation
of $c_0$. Once noticed that the most part of the PB failures are due to $c_2$ and
in a first attempt to understand the emergence of bursts in the network,
we can focus only on the firing times of neuron $c_0$ and $c_2$.

To get a deeper insight on this issue, let us consider the anti-resonances 
(corresponding to minima in the PB activity) observed during the SNS experiments performed on $c_0$ 
(see Fig.~\ref{impatto_di_a_suicritici} C). To interpret such minima we examined
the firing periods $T_0$ and $T_2$ of the neuron $c_0$ and $c_2$ once isolated from the network.
For the LIF model~\cite{LIFreview1} these are simply given by 
$T_0 = \tau_\mathrm{m} \ln[(I^{{\rm stim}}-V_{r})/(I^{{\rm stim}}-V_{th})]$
and $T_2 = \tau_\mathrm{m} \ln[(I^b_{c_2}-V_{r})/(I^b_{c_2}-V_{th})]$, where $I^{{\rm stim}}$ is the
stimulation current acting on $c_0$ and $I^b_{c_2}$ the intrinsic excitability of $c_2$.
As shown in Table~\ref{tab1} the PB minima were associated to rational ratios of these periods.
This amounts to exact frequency locking of the firing of the two neurons~\cite{pikov}, 
whenever this occurs the bursting activity is depressed or even suppressed. This because 
the build up of a burst relies on a precise temporal mismatch between the firing of neuron $c_0$
and $c_2$, which, in the case of exact locking, can be achieved quite rarely or even never.
Therefore, given the absence of any structural connection among these two neurons, the clique  
functionality relied on unsupervised coordinated firing of $c_0$ and $c_2$.

\begin{figure}[h!]
\begin{center}
\includegraphics*[angle=0,width=7.9cm]{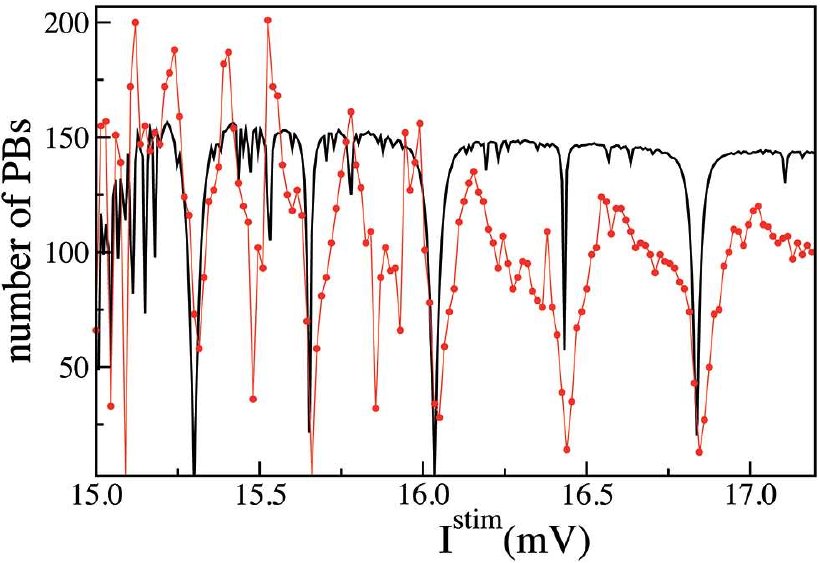}
\end{center}
\caption{ {\bf Model based reconstruction of the SNS experiment
for the critical neuron $c_0$.}  Number of emitted PBs as a function of 
the stimulation current $I^{\rm stim}$ applied to the neuron $c_0$.
The red line with dots refers to the results of
the SNS experiment on $c_0$ (same curve as in Fig.~\ref{impatto_di_a_suicritici} C)
and the black line to the estimations obtained by measuring the PB occurrence
with the simple model for SNS, described in the Methods. The measurement were
performed in both cases over a time interval $\Delta t = 84$ s.
}
\label{modello_periodi}
\end{figure}

In order to confirm this hypothesis, we developed a simple model to reproduce
the results of the SNS experiment on $c_0$. In particular, we assumed 
that $c_0$ and $c_2$ could be considered as two independently spiking neurons 
with their own firing periods determined by the stimulation current
$I_{stim}$ for $c_0$ and by the intrinsic excitability for $c_2$. Furthermore, we
assumed that a PB is emitted with a certain probability (related to the synaptic 
depression induced by the stimulation) whenever $c_0$ and $c_2$ fired in the correct order
and with a prescribed time delay (for more details see Methods). The results are
reported in Fig.~\ref{modello_periodi} and in the Table \ref{tab1}, the agreement is quite
surprising due to the limited ingredients employed in the model. Furthermore, the fact that more
than the 60\% of the ``anti-resonances'' as well as the level of the PB activity were reproduced
was a clear indication that the simple ingredients at the basis of the model represented the 
main mechanisms behind the PB build up process in this network. 
These mechanisms could be summarized as follows: the functional clique
can be assumed to be composed of two structurally connected pairs $c_0-c_1$ and $c_2-c_3$,
where $c_0$ and $c_2$ fired tonically and independently one from the other.
Any spike emission of $c_0$ induced a firing of $c_1$, however to recruit $c_3$ and 
therefore to initiate the PB, also $c_2$ should deliver a spike, with the right time delay
after $c_1$. Therefore, if $c_0$ and $c_2$ fired with periods which were
rational multiples one of the other it was unlikely to build up the PB. Since the
synchronism among the two neurons did not allow $c_1$ to participate to the build up
of the PB. The spike delivered by $c_1$ is fundamental to lead $c_3$ above threshold
and to trigger the avalanche, but it should be emitted at the right moment, 
as clearly shown in Fig.~\ref{PBrec} B and C.

\section*{Discussion}

The aim of the present work was to identify neuronal network arrangements sensitive 
to single neuron perturbations, such as those induced by single neuron stimulation
or deletion (or forced silencing). We choose as a benchmark model a random network 
of excitatory LIF neurons, connected via depressive synapses regulated by the TUM mechanism \cite{Tsodyks2000Synchrony}. 
Such networks displayed spontaneous bursting activity also in absence of inhibition, 
as extensively described in the literature~\cite{Tsodyks2000Synchrony,Loebel2002,Stetter2012,Divolo2013}. 
The choice of random topology was aimed at revealing the role of developmentally regulated neuronal excitability 
and connectivity gradients \cite{Bonifazi2009,Picardo2011,Marissal2012,Doetsch2005,Karayannis2012}, rather than 
specific topological configurations, in rendering network organization sensitive to single neuron perturbations. 

The introduction of a positive correlation between in- and out-degree (T1) and a negative correlation between 
intrinsic neuronal excitability and total degree (T2), besides being justified from a developmental point 
of view, favors also the stabilization of the network activity. This because, as pointed out in \cite{Vasquez2013},
in an excitatory network the sensitivity to fluctuations is mainly due to cells with a high out-degree.
Therefore, to avoid that their activation during spontaneous activity can cause network destabilization, 
a possible strategy is to impose an anti-correlation between their level of excitability and their degree, 
as done in the present work, or between in- and out-degree as shown in~\cite{Vasquez2013}.
Furthermore, when correlations T1 and T2 were embedded in the network, single neuron deletion/stimulation 
of a few peculiar neurons strongly impacted the frequency  of occurrence of population bursts. Most critical neurons, 
i.e. those capable of silencing the network when deleted or stimulated, shared common features: they constantly/reliably participated 
in the PB build up (i.e. they were functional hubs) and they had a quite low structural degree. 
These functional hubs formed a clique, where their precise ordered temporal activation was necessary for the burst generation. 
In the specific case here described, the clique was composed by two synaptically connected pairs, each composed of one neuron
above and one below threshold. The burst could be triggered only when the first three neurons operated
at precise time lags and the last neuron of the clique (which is just below threshold) was led to fire. 

Each population burst caused the depletion of the synaptic resources, 
therefore another PB could occur only when the synaptic resources would 
be recovered, thus inducing an effective refractory time between two successive PBs.
However, this is only a necessary, but not sufficient condition for PB triggering.
The key element responsible for generating PBs was the unsupervised occurrence of a precise sequence of 
firing times of the two supra-threshold critical neurons, i.e. not mediated by any structural synaptic connection. 
On the other hand, the mode locking of the firing frequencies of these two neurons 
was instead responsible for {\it anti-resonances} associated to a drastic reduction of the PBs.  
For random networks, i.e. with no correlations, or embedding just one of the correlations of type T1, T2, T3 or the
combination of type T1 and T3,  we did not find any evidence of functional
cliques and the mechanisms of network synchronization were much more robust and 
immune from single neuron perturbations (see Supplementary Information).  

The activity of random uncorrelated networks has been previously examined in ~\cite{Tsodyks2000Synchrony},
in particular the authors have shown that the elimination of a pool of neurons (namely, $30$ neurons, corresponding 
to the $\simeq 8$ \% of the excitatory population) led to the interruption of the bursting activity. 
The PBs were suppressed whenever the removed pool was composed by neurons with an intermediate firing 
rate $(\simeq 1.3-2.5$ Hz). These neurons were responsible for triggering the avalanches 
in the network, due to their effectively strong excitatory synapses
and to their proximity to the firing threshold. From these findings, it is clear that in an uncorrelated 
network the PBs emerge due to a cooperative effect involving a large portion of neurons. On the contrary, the 
introduction of correlations of type T1 and T2 induces single neuron sensitivity as discussed in this paper.

Furthermore, our results show that the integration into a clique is the key element that can enable 
single neurons to impact the population dynamics, without any further topological requirements 
for the network architecture. The functional hubs forming and operating within the clique, are actively
involved in generating network synchronizations and, as a consequence, capable
to impact the network dynamics when stimulated. Therefore, without necessarily being
structural or effective hubs, i.e. capable to cause a direct influence on the
activity of many other nodes \cite{friston1994,sporns2007}, they operate as {\it operational hubs} accordingly
to the definition recently introduced in~\cite{rosa2014}.
Similarly to the hub cells in the developing hippocampus whose stimulation was capable to 
drastically reduce the frequency of spontaneous network
synchronization~\cite{Bonifazi2009}, the critical neurons presented 
in this paper have a very high functional connectivity and several of them are close to the firing threshold. 

At variance with hippocampal hubs, critical neurons do not have a high structural degree. This is a consequence 
of the correlation imposed on the network where the excitability of the neurons is anti-correlated to the total 
structural degree of the cells. Indeed, in the correlated network 
studied in this work, the orchestration of the neuronal activity relies on the 
coordinated firing of a few critical ``young'' neurons (i.e. with a low structural degree)
mediated by their inter-connections. However, in real biological developing networks, it is possible that a further 
developmental connectivity regulation is fulfilled, with the chance of finding synaptic connections in a pair of young 
cells much lower compared to a pair composed of a young and a mature cells. This would be also in line 
to the {\it rich gets richer} rule which can generate scale-free networks \cite{Barabasi1999}. 
In such case, the orchestration between unconnected young neurons would require the presence of a 
structural connector or hub, i.e. a more developed neuron, capable to receive and promptly activate in 
the presence of a few synchronized inputs. 
Therefore, our study supports the hypothesis that, in developmentally constrained networks, 
PBs are triggered by a precise time activation of a few around threshold oscillators. This is
indeed the case of neurons $c_1$ and $c_3$, which are fundamental for the ignition of the neuronal avalanche,
but they need to be activated by a precise firing sequence involving $c_0$ and $c_2$.
This evidence is even more striking in the example discussed in Text S3 for $N=200$ neurons, 
where the functional clique is composed of a small group of neurons all just below threshold, 
apart the leader who activates the neurons in cascade leading to the burst.

We have verified that the main ingredients required to observe strong sensitivity to
single neuron stimulation and deletion are, besides the presence of the correlations
of type T1 and T2, a small number of neurons supra-threshold as well as a strongly diluted 
network. This can find an explanation in the fact that by increasing the degree of the neurons 
as well as the number of neurons supra-threshold the network dynamics becomes 
more cooperative. Furthermore, the synaptic time scales seem not to be crucial for 
the emergence of single neuron sensitivity (for more details see the subsection Dependence on the Model Parameters in Methods).

Although presented within a developmental framework, our results can also have elements of interest 
in the wider context of brain processing. In fact, in this work we show that the introduction of 
an excitability gradient in the network can lead to the emergence of functional cliques capable to 
shape the neuronal population activity. Indeed, different brain states could dynamically modulate the level 
of excitability or the gain function of the neurons within a circuit (as clearly 
discussed in \cite{haider2009}) and in this way instantaneously induce the emergence of functional cliques.
Furthermore, functional chains of neural activation have been reported also in the different framework 
of feed-forward networks~\cite{abeles1991,diesmann1999}. In this context,
in Ref.\cite{Jahnke2013} the authors found that structural hubs (i.e. highly connected neurons) have a peculiar role 
in promoting the signal transmission across sequences of non-hub sub-networks.

The previously discussed anti-resonance effect leading to the silence of the bursting activity
resembles recent results reported in literature~\cite{Lysyansky2011}, where the
authors have shown that abnormally synchronous neuronal populations can be desynchronized
by administrating stimulations at resonant frequencies to an ensemble of spiking neurons.
In that context, the desynchronization of the neuronal activity can be achieved by delivering a periodic 
stimulation at few sites, with a period which was an integer multiple of the fundamental period of the 
synchronized system. This is in striking contrast with what usually observed 
for a resonant forcing of a population of coupled oscillators~\cite{Rulkov2006,Kiss2008}. Our results, 
revealing population desynchronization associated to anti-resonances at the level of single neuron frequencies, are 
even more intriguing. On one side these findings suggest the possibility of extremely non-invasive procedures
to treat pathological neuronal synchronization, which is associated to several neurological disorders
\cite{Pare1990,Lenz1994}. On the other side they reveal the potentiality of a brain circuit
able to adapt to external stimuli on the basis of unsupervised mechanisms, which can switch the network
activity from coherent to incoherent.

The numerical results here presented predict a primary role for supra-threshold and near-threshold 
cells capable to impact network synchronizations when organized into functional clique. 
Probing the existence of such cliques, whose emergence could be also dynamically regulated by
varying the gradient of excitability in the circuits~\cite{haider2009}, is experimentally challenging, 
but surely feasible in {\it in-vitro} biological preparations.
Cultured networks allow for an easier access and probing of the circuits~\cite{Marissal2012,Bonifazi2013},
while representing a general model of unsupervised (or self-organized) spontaneous network synchronization
in circuits under development, analogously to what observed in central and peripheral brain circuits~\cite{marom2002,blankenship2010}. 
In these circuits high functionally connected neurons (mostly activated at the build up
of bursting) and highly active (i.e. supra-threshold) neurons
could be identified by using both multi-electrode recordings 
and/or calcium imaging~\cite{eytan2006,Bonifazi2013}.
Furthermore, by manipulating the frequency of firing of such cells through multi-site optical or
electrical stimulation~\cite{Marissal2012,Wallach2012} it is possible both to disrupt 
the sequential activation necessary for triggering network synchronizations 
(as displayed in Figs.~\ref{burst_synapse} and \ref{clique}) and 
to test the anti-resonance effects, as described in Fig.~\ref{modello_periodi} and Table~\ref{tab1}.
 
In this work, we considered a deterministic model of short term synaptic 
depression based on a trial-averaged representation. In recent papers, the stochastic 
processes involved in vesicle release and synaptic recovery time have been also taken
into account to model short-term synaptic depression~\cite{goldman2004,de2005,rosenbaum2012}.  
In particular, in Ref.~\cite{de2005} the authors compare
deterministic and stochastic model for short-term plasticity. 
They found that for supra-threshold neurons the two setups give essentially the same behavior, 
while for sub-threshold neurons, whose spiking activity is fluctuation driven,
the results of the deterministic and stochastic models essentially coincide 
for low frequencies (up to $\simeq 10$ Hz). In our study the functional hubs are found to be or
supra-threshold or to have a relatively low firing rate (see bottom panel in 
Fig. \ref{esempioesperimentoBonifazi} C). Therefore we expect that the implementation of 
stochastic short-term plasticity would not affect qualitatively our main findings, but
further investigations are required to fully clarify this issue.

\section*{Methods}
 
To study the response of bursting neural networks to single
neuron stimulation and removal, we employed the 
Tsodyks-Uziel-Markram (TUM) model~\cite{Tsodyks2000Synchrony}.
Despite being sufficiently simple to allow for extensive numerical
simulations and theoretical analysis, this model has been fruitfully utilized in 
neuroscience to interpret several phenomena~\cite{Loebel2002, Mongillo2008, Stetter2012}.
We have considered such a model, restricted to excitatory synapses, to somehow mimic 
the dynamics of developing brain circuitries, which is characterized by 
coherent bursting activities, such as {\it giant depolarizing potentials} \cite{Allene2008,Bonifazi2009}.
These coherent oscillations emerge, instead of abnormal synchronization,
despite the fact that the GABA transmitter has essentially an excitatory effect 
on immature neurons \cite{Ben2007}. 
The model uses leaky-integrate-and-fire (LIF) neurons with excitatory synapses
displaying short-term synaptic depression \cite{Tsodyks2000Synchrony} arranged
in a directed random network. 
It should be stressed that we do
not consider a network under topological development, which is typically
characterized by a dynamical
evolution (addition/deletion) of the links among the neurons.

\subsection*{The model}

In this paper we consider a network of $N$ excitatory LIF neurons,
interacting via synaptic currents regulated by short-term-plasticity
according to the model introduced in \cite{Tsodyks2000Synchrony}.
The time evolution of the membrane potential $V_i$ of each neuron reads as
\begin{equation}
\label{eq1n}
\tau_{\mathrm{m}} \dot V_i= -V_i + I^{\mathrm{syn}}_{i}+I^{\mathrm{b}}_{i} \, 
\end{equation}
where $\tau_\mathrm{m}$ is the membrane time constant, 
$I^{\mathrm{syn}}_{i}$ is the synaptic current received by neuron $i$ from 
all its presynaptic inputs and $I^{\mathrm{b}}_{i}$ represents its level of 
intrinsic excitability. The membrane input resistance is incorporated into the currents, 
which therefore are measured in voltage units (mV). 
   
Whenever the membrane potential $V_i(t)$ reaches the threshold 
value $V_{\mathrm{th}}$, it is reset to $V_{\mathrm{r}} $, and a spike is sent 
towards the postsynaptic neurons. For the sake of simplicity the spike is assumed to 
be a $\delta$--like function of time. 
Accordingly, the spike-train $S_j(t)$ produced by neuron $j$, is defined as,
\begin{equation}
S_j(t)=\sum_m \delta(t-t_{j}(m)),
\end{equation}
where $t_{j}(m)$ represent the $m$-th spike time emission of neuron $j$.
The transmission of the spike train $S_j$ to the efferent neurons is mediated by 
the synaptic evolution. In particular, by following \cite{Tsodyks1997} the state 
of the synaptic connection between the $j$th presynaptic neuron and the $i$th postsynaptic neuron is
described by three adimensional variables, $X_{ij}$, $Y_{ij}$, and $Z_{ij}$, 
which represent the fractions of synaptic transmitters in the recovered, active, and inactive state, 
respectively and which are linked by the constraint 
$X_{ij}+Y_{ij}+Z_{ij}=1$. The evolution equations for these variables read as
\begin{equation}
\label{dynsyn}
\dot Y_{ij} = -\frac{Y_{ij}}{T^I_{ij}} +u_{ij}X_{ij}S_{j}
\end{equation}
\begin{equation}
\label{contz}
\dot Z_{ij} = \frac{Y_{ij}}{T^\mathrm{I}_{ij}}  -   \frac{Z_{ij}}{T^\mathrm{R}_{ij}} .
\end{equation} 
Only the active transmitters react to the incoming spikes $S_j$: the adimensional parameters $u_{ij}$ 
tunes their effectiveness. Moreover, $\{T^\mathrm{I}_{ij}\}$ represent 
the characteristic decay times of the
postsynaptic current, while $\{T^\mathrm{R}_{ij}\}$ are the recovery times from synaptic depression. 

Finally, the synaptic current is expressed as the sum 
of all the active transmitters (post-synaptic currents) delivered to neuron $i$ 
\begin{equation}
\label{curr}
I^{\mathrm{syn}}_{i} = \frac{ G_{i}}{K^{I}_{i}}\sum_{j\ne i}  \epsilon_{ij}Y_{ij},
\end{equation}
where $G_{i}$ is the coupling strength, while $\epsilon_{ij}$ is the connectivity
matrix whose entries are set equal to $1$ ($0$) if the presynaptic neuron
$j$ is connected to (disconnected from) the postsynaptic neuron $i$. At variance 
with~\cite{Tsodyks2000Synchrony}, we assume that the coupling strengths are the same for all the
synapses afferent to a certain neuron $i$. We have verified that this simplification does not 
alter the main dynamical features of the TUM model under control conditions.
 
In this paper we study the case of excitatory coupling between neurons,
i.e. $G_{i} > 0$.  Moreover, we consider a {\it diluted} network made of $N=100$ neurons where the 
$i$-th neuron has $K^{I}_{i}$ ($K^{O}_{i}$) afferent (efferent) synaptic connections 
distributed as in a directed Erd\"os-R\'enyi graph  with average in-degree $\bar K^I =10$,
as a matter of fact also the average out-degree was $\bar K^0 =10$.
The sum appearing in (\ref{curr}) is normalized by the input degree $K^I_{i}$ 
to ensure homeostatic synaptic inputs~\cite{Turrigiano1998,Turrigiano2008}.

The propensity of neuron $i$ to transmit (receive) a spike can be measured in terms
of the average value of the fraction of the synaptic transmitters $X_i^{OUT}$ ($X_i^{IN}$)
in the recovered state associated to its efferent (afferent) synapses, namely
\begin{equation}
\label{XIN}
X_i^{OUT} = \frac{1}{K^O_i} \sum_{k \ne i} \epsilon_{ki} X_{ki} \quad, \hspace{0.4cm} 
X_i^{IN} = \frac{1}{K^I_i} \sum_{j \ne i}    \epsilon_{ij} X_{ij}   
\enskip .
\end{equation}

The intrinsic excitabilities of the single neurons $\{I^b_{i}\}$ are randomly chosen from a flat distribution
of width 0.45 mV centered around the value $V_{\mathrm{th}} = 15$ mV, with the constraint that
10\% of neurons are above threshold. This requirement was needed to obtain bursting behavior 
in the network. This choice ensures under control conditions that the distribution of the single
neuron firing rates is in the range $[0.03 ; 25]$ Hz.

For the other parameters, we use the following set of values: $\tau_\mathrm{m} = 30$ ms,  
$V_{\mathrm{r}} = 13.5$ mV, $V_{\mathrm{th}} = 15$ mV. The synaptic 
parameters $\{T^\mathrm{I}_{ij}\}$, $\{T^\mathrm{R}_{ij}\}$, $\{u_{ij}\}$ and $\{G_{i}\}$ are
Gaussian distributed with averages $\overline{T^\mathrm{I}} = 3$ ms, 
$\overline{T^\mathrm{R}} = 800$ ms, $\overline{u} = 0.5$ and $\overline{G} = 45$ mV, 
respectively, and with standard deviation equal to the half of the average.   
These parameter values are analogous to the ones employed in~\cite{Tsodyks2000Synchrony} and
have a phenomenological origin.

\subsection*{Correlations}

Furthermore, we have considered networks where correlations of type T1, T2 or T3 are embedded.
Correlation T1 is obtained by generating randomly two pools of $N-4$ input and output degrees 
from an Erd\"os-R\'enyi distribution with average degree equal to 10. The degrees are ordered
within each pool and then assigned to $N-4$ neurons in order to obtain a positive correlation between $K^{O}_{i}$ and $K^{I}_{i}$.
Finally, four hubs with total degree $K^T > 50$ are added to this $N-4$ neurons. The final total degree
distribution is shown in Fig. S1B.

Correlation of type T2 (T3) imposes a negative (positive) correlation between
excitability $I^b_{i}$ and the total degree of the single neuron $K^T_i =K^{I}_{i} + K^{O}_{i}$.
To generate this kind of correlation the intrinsic excitabilities are randomly generated,
as explained above, and then assigned to the various neurons 
accordingly to their total connectivities $K_i^T$, thus to ensure an inverse (direct) correlation 
between $I^b_{i}$ and $K_i^T$. Correlations of type T2 (T3) are visualized in 
Fig. S1 A and Fig. S6 A.

\subsection*{Numerical Integration of the Model}

In order to have an accurate and fast integration scheme, we transformed
the set of ordinary differential equations (\ref{eq1n}), (\ref{dynsyn}) and (\ref{contz}) 
into an event--driven map \cite{Zillmer2007} ruling the evolution of the network
from a spike emission to the next one (see Text S1 for more details on the implementation
of the event--driven map). It is worth to stress that the event--driven formulation is an exact 
rewriting of the dynamical evolution and that it does not involve any approximation.

\subsection*{Population Bursts}

In order to identify a population burst we have binned the spiking activity 
of the network in time windows of 10 ms. A population burst is identified whenever the spike 
count involves more than 25 \% of  the neural population. 
In order to study the PB build up, a higher temporal resolution 
was needed and the spiking activity was binned in time windows of 1 ms. 
The peak of the activation was used as time origin (or center of the PB) and 
it was characterized by more than 5\% of the neurons firing within a 1 ms bin. 
The time window of 25 ms preceding the peak of the PB was considered as the build up period
for the burst. In particular, the threshold crossing times have been defined via a simple 
linear interpolation based on the spike counts measured in successive time bins.

These PB definitions gave consistent results for 
all the studied properties of the network. The employed  burst detection
procedure did not depend significantly on the precise choice of the threshold, since during
the inter-burst periods (lasting hundreds of milliseconds) only 10 - 15 \% of neurons were
typically firing, while more than 80 \% of the neuronal population contributed 
to the bursting event (lasting $\simeq 30$ ms).

The average interburst interval for the network with (without) correlations under
control conditions was $586\pm183$ ms ($208\pm74$ ms) for a network made of
$N=100$ neurons, while the burst duration was $27\pm 3$ ms.
Doubling the number of neurons in the correlated network
did not affect particularly neither the average interburst, which became $\simeq 500$ ms,
nor the burst duration ($\simeq 24$ ms). 
For a more detailed discussion of the dynamics of this larger network 
see the Text S3.

\subsection*{Dependence on the Model Parameters}

In this subsection, we summarize the crucial ingredients needed to observe strong
sensitivity to SNS/SND. In particular, we have checked, for different model parameters, 
when SNS/SND experiments were still able to noticeably modify the PB activity 
(i.e. more than 90 \% with respect to control conditions). 
Firstly, we considered the sparseness of the network, the reported
results refer to a diluted network with a probability of 10 \% to have a link 
among two neurons. We have observed that the results of the SNS/SND experiments strongly 
depend on the level of dilution, however they can still be effective up to a connection 
probability of 50 \%. Another crucial aspect was the small number of neurons supra-threshold,
in the studied case this number corresponded to the 10 \% of the neurons. By varying
this percentage up to 20 \% we still observed that the network can be silenced by
single neuron stimulation/removal. Furthermore, the dependence on the system size seemed
not be crucial, since as described in Text S3
by doubling the system size a functional clique can still be identified.

We also tested the influence of the synaptic time constants on the population sensitivity to SND/SNS.
As a matter of fact, the time scale ruling the depletion of the neurotransmitter $\overline{T^\mathrm{I}}$ 
affects the duration of the PB, while the recovery time from the synaptic depression $\overline{T^\mathrm{R}}$
influences the intervals between consecutive PBs~\cite{Tsodyks2000Synchrony}.
By varying $\overline{T^\mathrm{I}}$ within the interval $[1.5;4.5]$ ms, while
keeping fixed the ratio with $\overline{T^\mathrm{R}}$, we do not observe substantial modifications on the
network dynamics. Furthermore we found strong response to SNS and SND, leading to the population
silence for several stimulated neurons,  both for faster and slower synaptic time scale
than the one actually employed in the studied example. However, it should be noticed that
by increasing $\overline{T^\mathrm{I}}$ to extremely large values (namely, $\overline{T^\mathrm{I}} > 20-30$ ms)
this destroys the bursting behaviour in the network, which is then substituted by an asynchronous
activity \cite{Divolo2013}.

\subsection*{Functional Connectivity}

In order to highlight statistically significant time-lagged activations of neurons, 
for every possible neuronal pair 
we measured the cross-correlation between their spike time series. On the basis of 
this cross-correlation we eventually assign a directed functional 
connection among the two considered neurons, similarly to what reported 
in \cite{Bonifazi2009, Bonifazi2013} for calcium imaging studies.

Let us explain how we proceeded in more details. 
For every neuron, the action potentials timestamps were first converted into a binary 
time series with one millisecond time resolution, where ones (zeros) marked the occurrence 
(absence) of the action potentials. Given the binary time series of two neurons $a$ and $b$, the 
cross correlation was then calculated as follows:
\begin{equation}
C_{ab} (\tau) = \frac{\sum_{t=\tau}^{T-\tau} a_{t+\tau} b_t}{min(\sum_{i=1}^T a_i, \sum_{k=1}^T b_k)} 
\end{equation}
where $\{a_t\}$,$\{b_t\}$ represented the considered time series and $T$ was their total duration.
Whenever $C_{ab} (\tau)$ presented a maximum at some finite time value $\tau_{max}$ a functional
connection was assigned between the two neurons: for $\tau_{max} < 0$ ($\tau_{max} > 0$) 
directed from $a$ to $b$ (from $b$ to $a$). A directed functional connection cannot be defined
for an uniform cross-correlation corresponding to uncorrelated neurons or for synchronous firing
of the two neurons associated to a Gaussian $C_{ab} (\tau)$ centered at zero.
To exclude the possibility that the cross correlation could be described by a Gaussian with zero mean 
or by a uniform distribution we employed both the Student\'\-s t-test and the Kolmogorov-Smirnov test with 
a level of confidence of 5\%. The functional out-degree $D^O_i$ (in-degree $D^I_i$) of a neuron $i$ 
corresponded to the number of neurons which were reliably activated after (before) its firing.

\subsubsection*{Time series surrogates}

In order to treat as an unique event multiple spike emissions occurring within a PB, 
different time series surrogates were defined for different kind of analysis according to 
the following procedures:

\begin{enumerate}

\item for the definition of the functional in-degree $D^I_i$ and out-degree $D^O_i$, 
all the spiking events associated to an inter-spike interval longer
than $35$ ms were considered. Since we observed that this was the minimal duration 
of an inter-spike outside a PB and it was larger than the average duration of the PBs. 
This implies that for each neuron only the timestamp of the first spike within a PB was kept;

\item for the description of the PBs build up only the timestamps of the first action 
potential emitted within a window of $25$ ms preceding the PB peak was taken into account;

\item for the analysis of the network activity during inter-burst periods, all action potentials 
emitted out of the PBs were considered.

\end{enumerate}

\subsection*{A simple model for SNS}

The model here reported has been developed to reproduce the network response
during the SNS experiments on $c_0$ for a range of stimulation current $I^{{\rm stim}}$, 
which is displayed in Fig.~\ref{impatto_di_a_suicritici} C. To mimic this activity,
we only considered the dynamics of the two neurons of the 
clique $c_0$ and $c_2$ which were supra-threshold. 
In particular, we assumed that these two neurons fired tonically
and independently as they would be two isolated LIF neurons (oscillators)~\cite{LIFreview1}.
Therefore, as a first step we generated two regular spike trains, one for $c_0$ and one for $c_2$ 
with constant inter-spike time intervals $T_0 = \tau_\mathrm{m} \ln[(I^{{\rm stim}}-V_{r})/(I^{{\rm stim}}-V_{th})]$
and $T_2 = \tau_\mathrm{m} \ln[(I^b_{c_2}-V_{r})/(I^b_{c_2}-V_{th})]$, respectively.
Successively, by examining the two spike trains, we assumed that a PB in the network could occur 
whenever the neuron $c_2$ fired after $c_0$ with a certain time delay $\tau_D$. Furthermore, we 
also assumed that the PB emission was a probabilistic event with a 
finite probability ${\cal P}= {\cal P}(I^{{\rm stim}})$.
${\cal P}$ is simply given by the average efferent synaptic strength $\langle X^{OUT}_{c_0} \rangle$ 
measured during the SNS experiment on $c_0$ suitably 
rescaled in order to get probability one for $I^{{\rm stim}} = V_{th}$.
This probability has been introduced to mimic the decrease of the effective synaptic strength
induced by the increasing stimulation, as shown in Fig.~\ref{effective_synapse} C.

In summary, for each stimulation current $I^{{\rm stim}}$ we considered the sequence of the 
firing times of $c_0$ and $c_2$ and we registered the occurrence of a PB whenever the two following 
conditions were both fulfilled

\begin{itemize}

\item $c_2$ fired after $c_0$  within a time window $[\tau_D - \sigma_D;\tau_D + \sigma_D]$,
where $\tau_D = 14 $ ms is the average time delay $\Delta T_{c_2,c_0}$ measured immediately before
a population burst and $\sigma_D = 4$ ms is its standard deviation, both measured
in control conditions;

\item a random number $r$ extracted by a flat random distribution with support $[0:1]$
was smaller than ${\cal P}(I^{{\rm stim}})$.

\end{itemize}

Furthermore, each time a PB was registered, for the subsequent $27 \pm 3$ ms
(corresponding to the duration of a PB under control conditions)
no further PB could be counted. The results of this simple model are reported in Fig.~\ref{modello_periodi}
and Table \ref{tab1} together with the numerical results obtained by the simulation of the network activity.

\section*{Acknowledgments}

We acknowledge useful discussions with D. Angulo Garcia,  N. Bozanic, J. Challenger,
T. Kreuz, A. Politi, O.V. Popovych, G. Puccioni foundation for the project G.I.B.B.O
and F. Cherubini for the technical help in the realization of the figures.


\clearpage

\section*{Tables}

\begin{table}[!ht]
\caption{\bf{Anti-resonances observed during SNS of $c_0$}}
\begin{center}
\begin{tabular}{|c|c||c|}
\hline
\hline
& & \\
$I^{{\rm stim}}$ (mV) &  $T_0/T_2$ &  $I^{{\rm stim}}$ (mV) (Model) \\
& & \\
\hline
\hline
15.04 & 2 &  15.04 \\
15.09  &  8/5 & - \\ 
15.31   &   1   & 15.30 \\
15.48  &   4/5  & 15.53 \\
15.66 &  2/3   &  15.65 \\
15.88 &   0.565 &  - \\
16.05  &  1/2 &  16.03 \\
16.44  &  2/5  & 16.43 \\
16.84 & 1/3 & 16.83 \\
17.28 &  0.282 & - \\
17.73 &  1/4 & 17.65 \\
\hline
\hline
\end{tabular}
\end{center}
The first column reports the stimulation currents $I^{{\rm stim}}$ for which pronounced minima (anti-resonances)
are observed in the stimulated PB activity during SNS experiment
on $c_0$ (same data as in Fig.~\ref{impatto_di_a_suicritici} C and red curve in Fig.~\ref{modello_periodi}),
the second column the corresponding $T_0/T_2$ ratios. $T_0$ and $T_2$
are the firing periods of the LIF neurons $c_0$ and $c_2$ in isolation,
namely, $T_0 = \tau_\mathrm{m} \ln[(I^{{\rm stim}}-V_{r})/(I^{{\rm stim}}-V_{th})]$
and $T_2 = \tau_\mathrm{m} \ln[(I^b_{c_2}-V_{r})/(I^b_{c_2}-V_{th})]$.
The third column refers to the anti-resonances generated by employing
the simple model for SNS of $c_0$ introduced in the Methods (same data 
as the black curve in Fig.~\ref{modello_periodi}). The reported values 
correspond to the minima in the PB activity for this model, the absence of
a value means that the model did not display a corresponding minimum. 
The data refer to SNS experiments performed 
over a time interval of duration $84$ s.
\label{tab1}
\end{table}

\clearpage


\clearpage

\clearpage

\section*{Supporting Information Legends}

\begin{center}
\includegraphics*[angle=0,width=16.0cm]{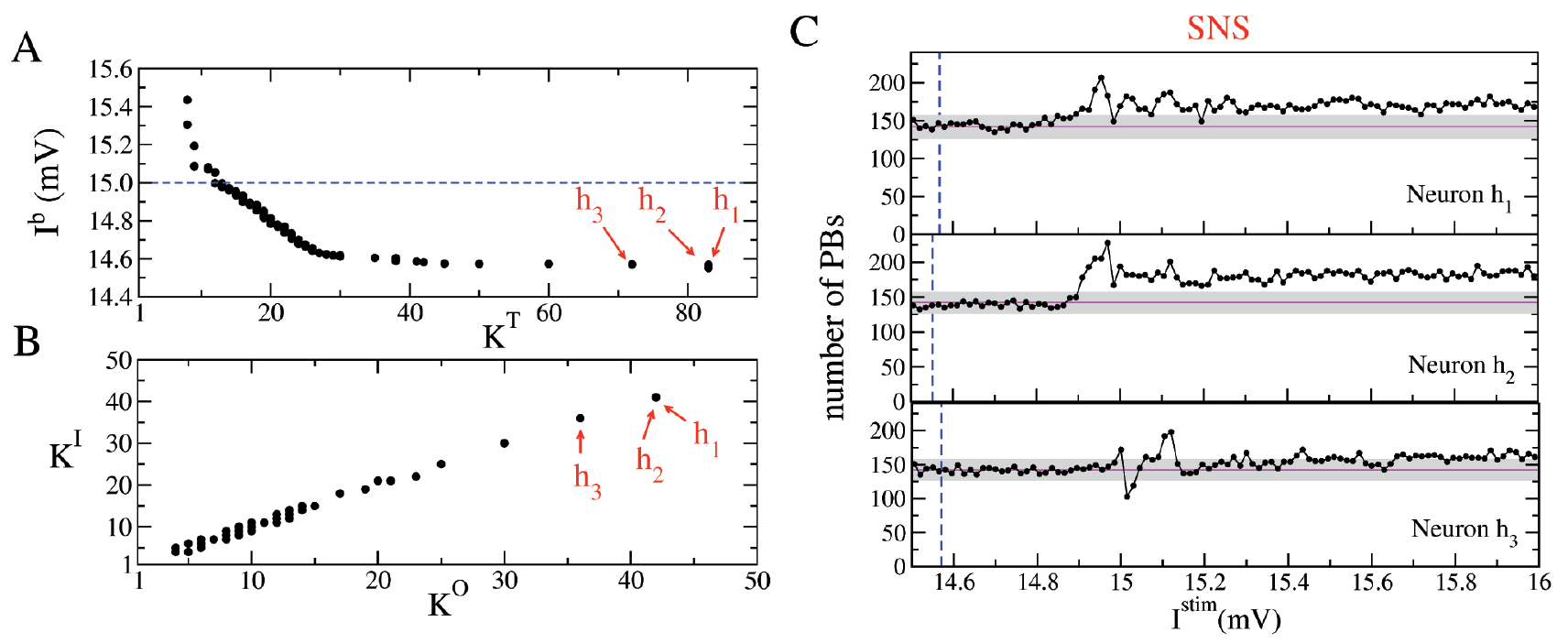}
\end{center}

\subsection*{Legend of Figure S1}

{\bf Network of $N=100$ neurons with negative correlation between $I^b$ and $K^T$ and
positive correlation between $K^I$ and $K^O$ (Setup T1 plus T2).}
(A) Negative correlation between intrinsic excitability $I^b$ and total connectivity $K^{T}$. 
The blue dashed line indicates the threshold value $V_{th} = 15$ mV.
(B) Positive correlation between the in-degree $K^I$ and the out-degree $K^O$.
All the parameter values are defined as in Methods. The red arrows signal the neurons with 
the highest degrees $h_1$, $h_2$, $h_3$. (C) Number of PBs emitted during SNS of 
the neurons $h_1$, $h_2$, $h_3$ versus the stimulation current $I^{stim}$. 
The blue vertical dashed lines mark the value of the intrinsic excitability in control
condition, while the magenta horizontal solid lines indicate average number of emitted
bursts under control conditions and the shaded grey area denote 
the amplitude of the fluctuations (measured as three standard deviations). 
The number of PBs are measured in all reported experiments 
over a time interval $\Delta t=$ 84 s.

\clearpage

\begin{center}
\includegraphics*[angle=0,width=16.0cm]{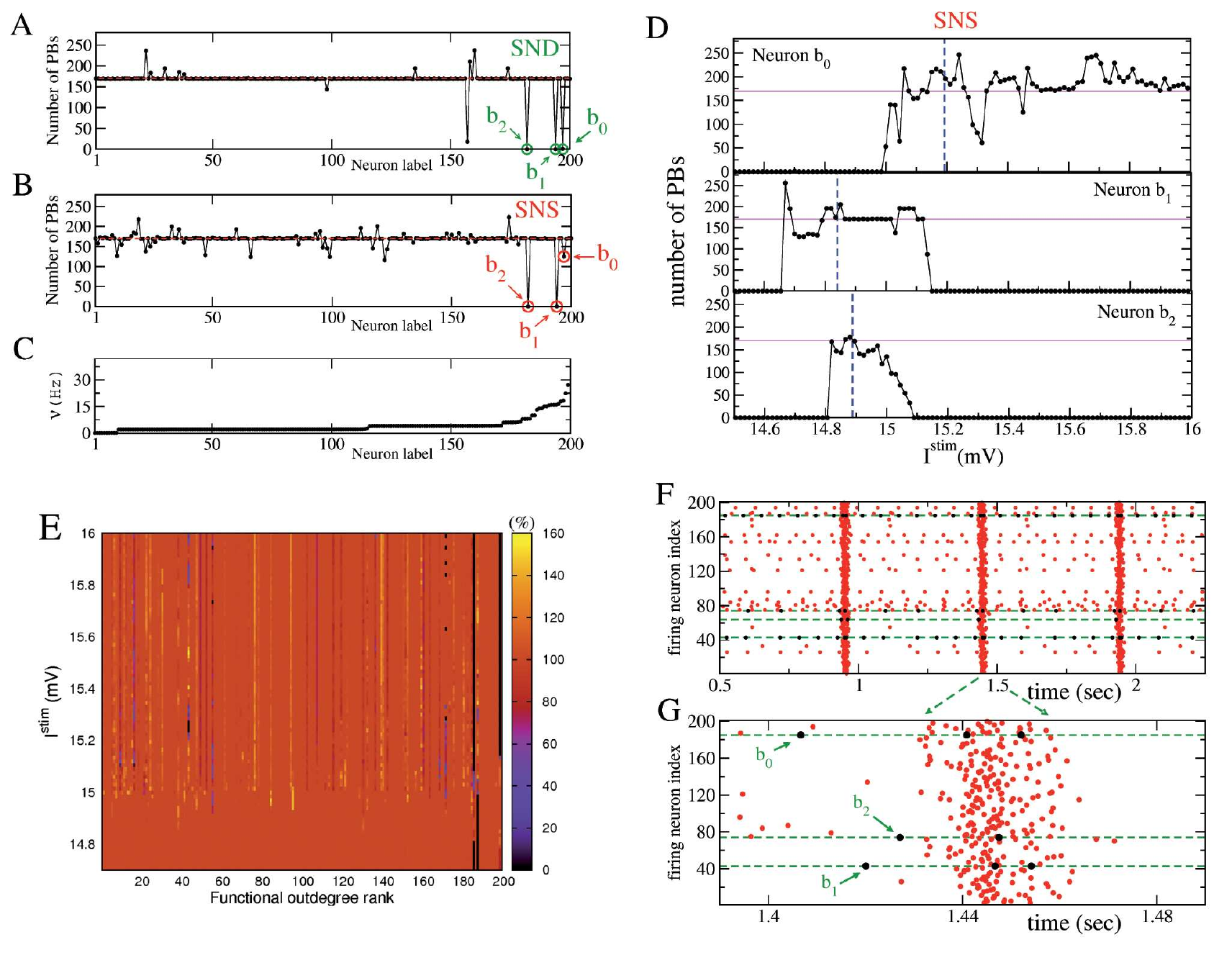}
\end{center}
 
\subsection*{Legend of Figure S2}

{\bf Network of $N=200$ neurons with negative
correlation between $I^b$ and $K^T$ and
positive correlation between $K^I$ and $K^O$ (Setup T1 plus T2).} 
(A), (B) Number of PBs emitted in a time window $\Delta t=84$ s during 
single neuron deletion (SND) experiments (A) and
single neuron stimulation (SNS) experiments (B) with $I^\mathrm{stim}=15.45$ mV. 
The horizontal dashed lines refer to the average number of PBs emitted
in a time interval $\Delta t=84$ s during a control experiment when no stimulation is applied 
(the amplitude of the fluctuations is smaller than the symbols). 
Neurons are ordered accordingly to their average firing rates $\nu$ as measured during 
control condition (data shown in panel (C)). 
In the figure the green (red) circles mark the critical 
neurons $b_0$, $b_1$, $b_2$, which under SND (SNS) can strongly 
affect the bursting activity of the network. 
(D) Impact on the network dynamics due to SNS of the critical neurons $b_0$, $b_1$, $b_2$ with 
various stimulation currents in the interval $I^\mathrm{stim} \in [14.5:16.0]$ mV. 
The blue vertical dashed lines and the magenta horizontal solid lines mark, resp., 
the value of the intrinsic excitability and the bursting activity of the network during control conditions. 
The number of PBs are measured over a time interval $\Delta t=$ 84 s. (E) Color coded rates of emission of PBs 
during SNS experiment performed for a range of injected DC currents $I^{stim}$ (y-axis).
The PB rates during SNS are normalized to the PB rate in resting conditions. 
Neurons are ordered according to their functional out-degree rank (x-axis). 
The number of PBs are measured over a time interval $\Delta t=84$ s. (F) Raster plot of the 
network activity: every dot denotes a firing event. The (green) dashed lines and (black) dots refer to the 
critical neurons. (G) Enlargement of a representative population burst: PBs are anticipated by the 
ordered firing sequence $b_0 \to b_1 \to b_2$. For clarity reasons, in the raster plots, 
at variance with all the other figures, the neuronal labels are not ordered 
accordingly to their firing rates.

\clearpage

\begin{center}
\includegraphics*[angle=0,width=16.0cm]{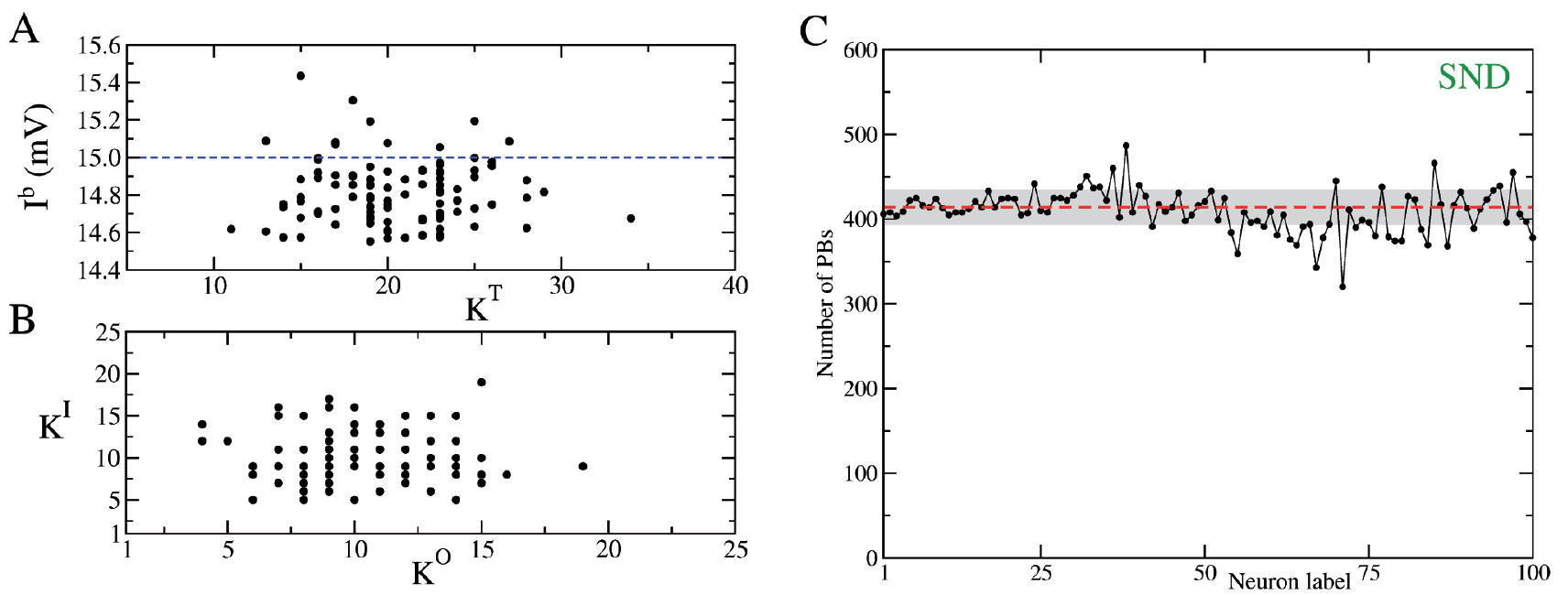}
\end{center}

\subsection*{Legend of Figure S3}

{\bf Network of $N=100$ without any correlation.}
(A) Distribution of the intrinsic excitabilities $I^b$ versus the corresponding 
total degrees $K^T$. (B) Distribution of the in-degrees  $K^I$ versus out-degrees $K^O$ for 
each neuron in the network. (C) Number of population bursts, PBs, computed over the 
time interval $\Delta t$ = 84 s in 
simulations where neurons are one by one removed by the network (SND). 
The horizontal dashed line refers to the average number of PBs emitted 
within the same time interval during the control experiment, 
while the shaded grey area around the line denotes 
the amplitude of the fluctuations (measured as three standard deviations).
Here and in the following figures the data refer to $N=100$ and all the 
parameter values are reported in Methods.

\clearpage

\begin{center}
\includegraphics*[angle=0,width=16.0cm]{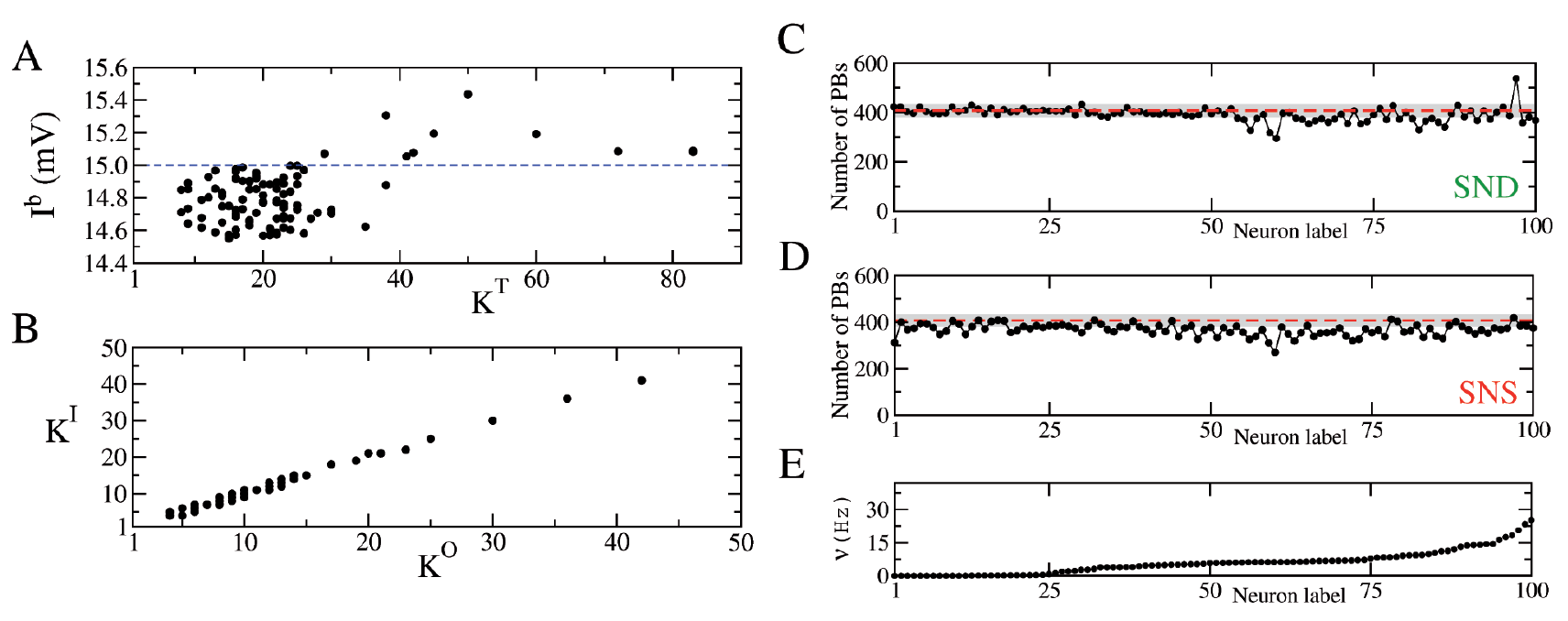}
\end{center}

\subsection*{Legend of Figure S4}

{\bf Network with positive correlation between in-degree and out-degree (Setup T1).}
(A) Distribution of the intrinsic excitabilities $I^b$ versus the corresponding 
total degrees $K^T$. (B) Distribution of the in-degrees  $K^I$ versus out-degrees $K^O$ for 
each neuron in the network. (C) Number of population bursts, PBs, measured during SND experiments where 
neurons are one by one taken out from the network. (D) PBs emitted during SNS with a stimulation
current $I^{{\rm stim}}=15.9$ mV.(E) Single neuron frequencies $\nu$ measured in a control experiment. 
In (C) and (D) the horizontal dashed lines and the shaded grey areas around the lines have the same meaning as in Fig.~S2. 
All the reported data have been measured over a time window $\Delta t$ = 84 s.

\clearpage

\begin{center}
\includegraphics*[angle=0,width=16.0cm]{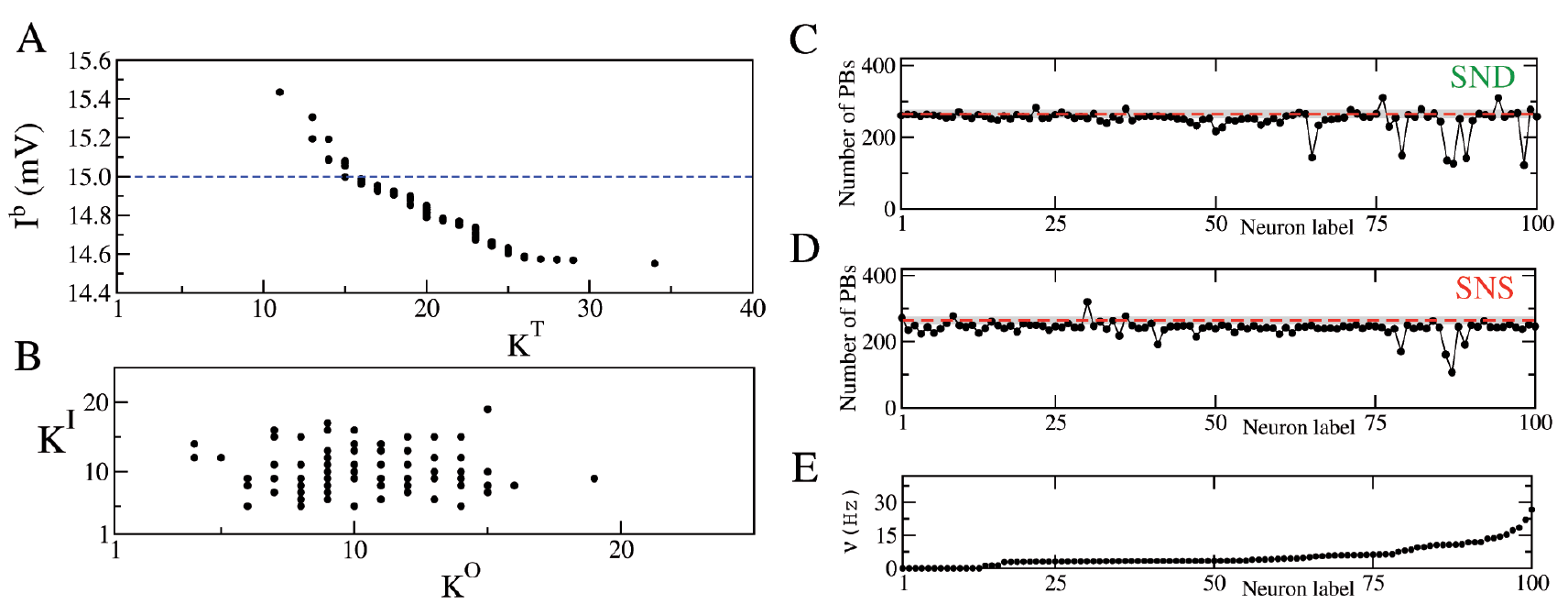}
\end{center}

\subsection*{Legend of Figure S5}

{\bf Network with anticorrelation between intrinsic excitability and total degree (Setup T2).}
(A) Distribution of the intrinsic excitabilities $I^b$ versus the corresponding 
total degrees $K^T$. (B) Distribution of the in-degrees  $K^I$ versus out-degrees $K^O$ for 
each neuron in the network. (C) Number of population bursts, PBs, measured during SND experiments 
where neurons are one by one removed from the network. (D) PBs emitted during SNS with a stimulation
current $I^{{\rm stim}}=15.9$ mV. (E) Single neuron frequencies $\nu$ as measured during the 
control experiment. In (C) and (D) the horizontal dashed lines and the shaded grey 
areas around the lines have the same meaning as in Fig.~S2. All the reported data have been measured over a time window $\Delta t$ = 84 s.

\clearpage

\begin{center}
\includegraphics*[angle=0,width=16.0cm]{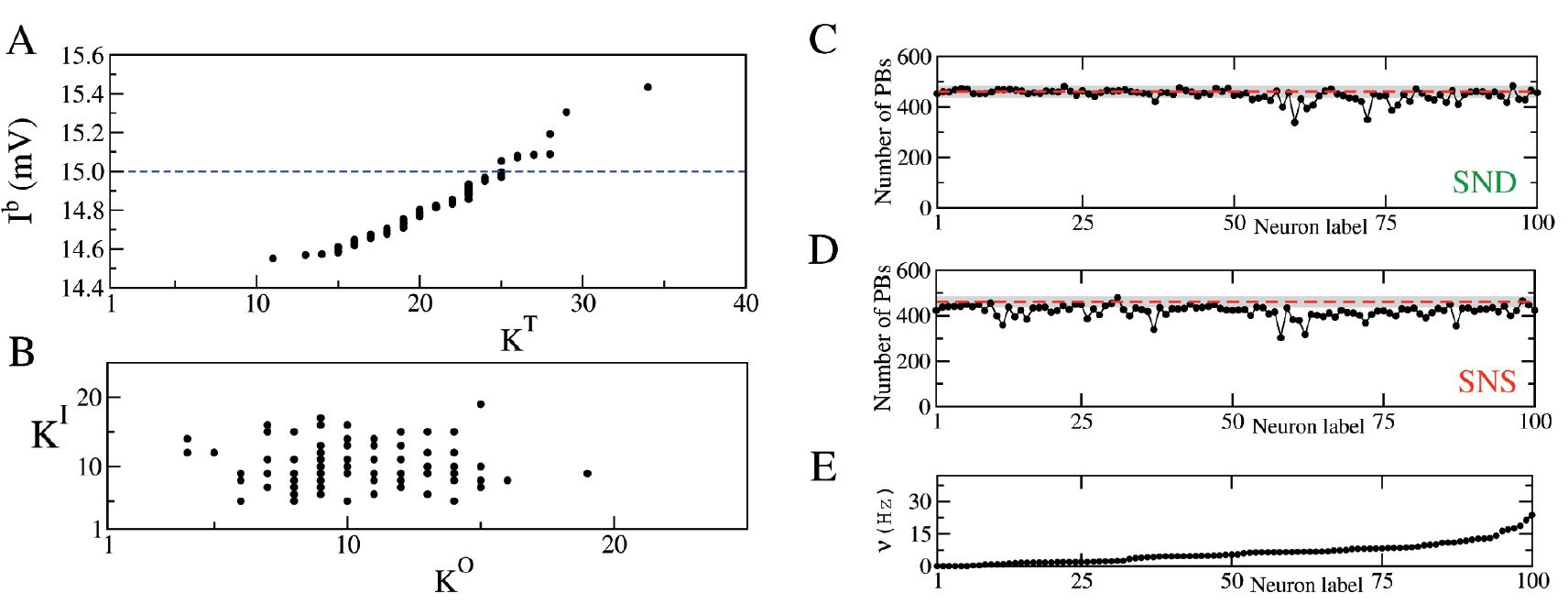}
\end{center}

\subsection*{Legend of Figure S6}

{\bf Network with positive correlation between excitability and total degree (Setup T3).} 
(A) Distribution of the intrinsic excitabilities $I^b$ versus the corresponding 
total degrees $K^T$. (B) Distribution of the in-degrees  $K^I$ versus out-degrees $K^O$ for 
each neuron in the network. (C) Number of population bursts, PBs, emitted during SND experiments 
where neurons are removed from the network one by one. (D) PBs emitted during SNS with a 
stimulation current $I^{{\rm stim}}=15.9$ mV. (E) Single neuron frequencies $\nu$ measured during a control experiment. 
In (C) and (D) the horizontal dashed lines and the shaded grey areas around the lines have 
the same meaning as in Fig.~S2. All the reported data have been measured over a time window $\Delta t$ = 84 s.

\clearpage

\begin{center}
\includegraphics*[angle=0,width=16.0cm]{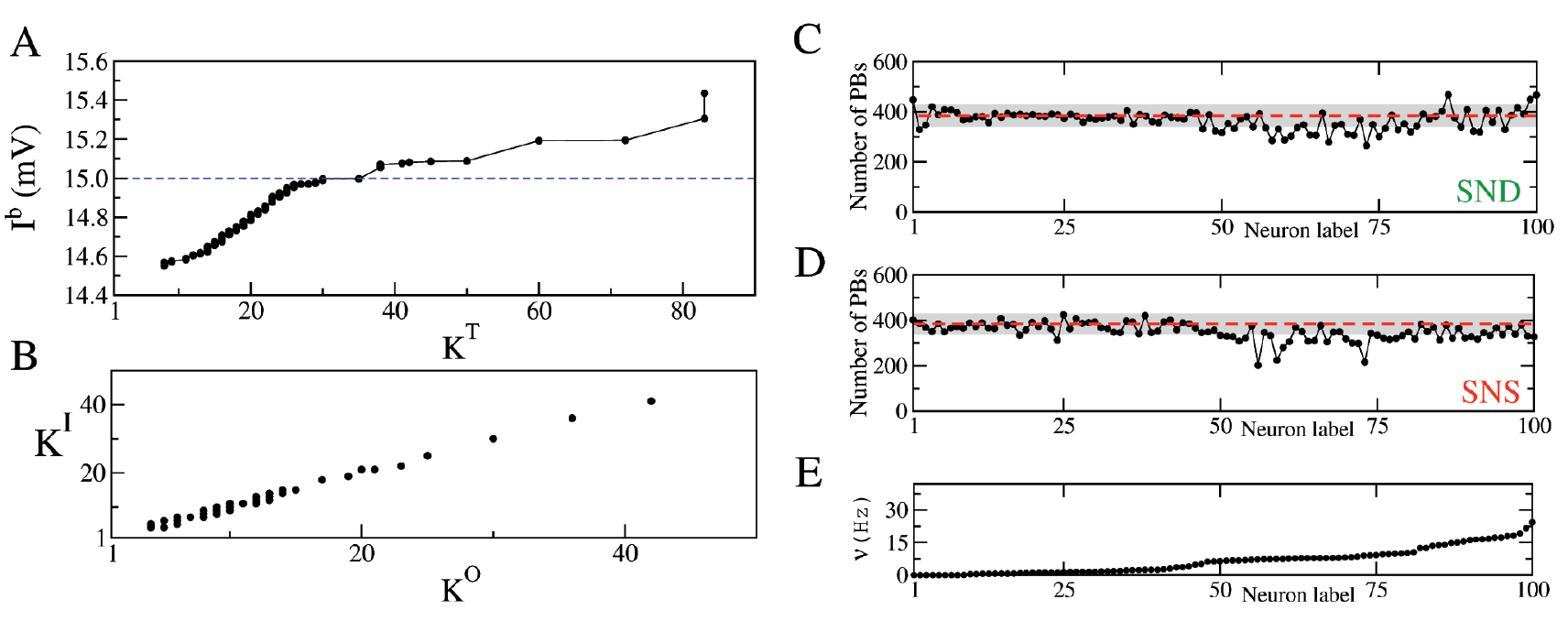}
\end{center}

\subsection*{Legend of Figure S7}

{\bf Network with positive correlation between intrinsic
excitability and total degree and 
positive correlation between in-degree and out-degree (Setup T1 plus T3).} 
(A) Distribution of the intrinsic excitabilities $I^b$ versus the corresponding 
total degrees $K^T$. (B) Distribution of the in-degrees  $K^I$ versus out-degrees $K^O$ for 
each neuron in the network. (C) Number of population bursts, PBs, measured during SND experiments 
where neurons are one by one removed from the network. (D) PBs emitted during SNS with a 
stimulation current $I^{{\rm stim}}=15.9$ mV. 
(E) Single neuron frequencies $\nu$ measured in control conditions. 
In (C) and (D) the horizontal dashed lines and the shaded grey areas around the lines have the same meaning as in Fig.~S2.
 All the reported data have been measured over a time window $\Delta t$ = 84 s.

\clearpage

\subsection*{Text S1: Event driven map}

A major advantage for numerical simulations comes from the possibility of
transforming the set of differential equations (1), (3) and (4) appearing in 
the article into an event--driven map \cite{Zillmer2007}. In fact, these differential
equations can be formally integrated from time $t_\mathrm{n}$ to time $t_{\mathrm{n+1}}$, where
$t_{\mathrm{n}}$ is the instant of time immediately after the emission of the $n$-th spike in
the network, to obtain a discrete time evolution from one spike to the successive one.
The resulting map for neuron $i$ reads
\begin{align}
\label{mapv}
& \begin{aligned}
& V_i(n+1)=V_i(n)e^{-\frac{\tau(n)}{\tau_m}} +I^{b}_{i}\Big(1-e^{-\frac{\tau(n)}{\tau_m}}\Big) +G_i F_{i}(n)
\end{aligned}\\
\label{mapz}
& \begin{aligned}
Z_{ij}(n+1)&=Z_{ij}(n)e^{-\frac{\tau(n)}{T^R_{\mathrm{ij}}}} +\frac{T^R_{\mathrm{ij}}}
{T^R_{\mathrm{ij}}-T^I_{\mathrm{ij}}}Y_{ij}(n)\Big(
e^{-\frac{\tau(n)}{T^R_{\mathrm{ij}}}} 
-e^{-\frac{\tau(n)}{T^I_{\mathrm{ij}}}}\Big)
\end{aligned}\\
\label{mapy}
& \begin{aligned}
Y_{ij}(n+1) & =Y_{ij}(n)e^{-\frac{\tau(n)}{T^I_{\mathrm{ij}}}}
+ u_{ij} \Bigg[1-\frac{T^R_\mathrm{ij}}{T^R_\mathrm{ij}-T^I_{\mathrm{ij}}}Y_{ij}(n)
\Bigg(e^{-\tfrac{\tau(n)}{T^R_\mathrm{ij}}}-\frac{T^I_{\mathrm{ij}}e^{-\frac{\tau(n)}
{T^I_{\mathrm{ij}}}}}{T^R_\mathrm{ij}}\Bigg)
-Z_{ij}(n)e^{-\frac{\tau(n)}{T^R_\mathrm{ij}}} 
\Bigg]\delta_{i,s},
\end{aligned}
\end{align}
where the index $s$ refers to the neuron spiking at time $t_{n+1}$,
$\tau(n)=t_{n+1}-t_n$ is the $n$-th inter--spike--interval (ISI) in the
network and
$F_i(n)$ has the following expression,
\begin{equation}
\label{eq2}
F_{i}(n)=\frac{1}{K^I_i}\sum_{j\ne i} \epsilon_{ij} Y_{ij}(n)\frac{T^I_{\mathrm{ij}}}{T^I_{\mathrm{ij}}-1}
\Big(e^{-\frac{\tau(n)}{T^I_{\mathrm{ij}}}}-e^{-\frac{\tau(n)}{\tau_m}}\Big),
\end{equation}
with the sum running over the index $j$, which denotes all direct connections
(afferent synapses) reaching neuron $i$ from all the other neurons. 
Notice that $\tau(n)$ can be determined by computing the time
\begin{equation}
\label{eq3}
\tau_i(n)=\mathrm{ln}\Bigg[\frac{I^b_i-V_i(n)}{I^b_i+G_i F_i(n)-1}\Bigg], \,\, i=1,
\cdots N \, ;
\end{equation}
needed to each neuron $i$th neuron to reach the threshold value and by selecting
the shortest one, namely
$$\tau(n)=\inf_{i} \{\tau_i(n) | i = 1, 2, \cdots , N\}.$$

\clearpage

\subsection*{Text S2: Dependence on different network realizations}
 
We repeated the SNS/SND experiments for eight different 
realizations at $N=100$
of the intrinsic excitability and of the synaptic parameters as well as 
of the connectivity matrix. In all the performed experiments
the parameter values are taken from  
random distributions with the same averages and standard deviations
as reported in Methods. Furthermore, the networks have been realized with the 
same average in-degree and the same constraints described in Methods.
In five of the examined cases we found a response to SNS/SND
experiments qualitatively similar to the one examined in the paper. 
In particular, we have identified a number of critical neurons ranging
from 6 to 3 in each realization with associated intrinsic excitability in 
the range $[14.82:15.30]$ mV, i.e. slightly below or above the firing threshold.
In most of the cases the critical clique contained two neurons supra-threshold,
as in the example discussed in the paper.
Furthermore, in four cases the stimulation/deletion of the critical neurons lead to the total 
suppression of the bursting activity (within the considered time interval), while in one case 
we observed very strong reductions of the population bursts (up to 65\% of the activity recorded 
in control conditions).

\clearpage

\subsection*{Text S3: Dependence on the network size}

In order to test if the size had any influence in the observed
phenomena, we considered a directed random graph  with
$N = 200$ neurons and an average in-degree $\bar K^I =10$ 
and with the embedded correlations of type T1 and T2. 
The network activity was still characterized by bursts of duration $\simeq 24$ ms
and with interburst intervals $\simeq 500$ ms, thus suggesting that the 
doubling of the size had not altered the main dynamical features of the network.
 
We performed SND and SNS experiments also in the present case.
SND experiments revealed that the removal of 3 critical neurons can lead
to the complete silence of the network, these neurons are identified in Fig.~S2 by
the labels $b_0$, $b_1$ and $b_2$. 
As shown in Fig.~S2 B, the SNS experiments, performed 
with stimulation current $I^\mathrm{stim}=15.45$ mV, confirmed the capability of 
neurons $b_1$ and $b_2$ 
to silence the network, while SNS on neuron $b_0$ reduced the PBs of $\simeq 27\%$.
The most critical neurons were all unable to fire if isolated from the network,
i.e. they were all below threshold, apart neuron $b_0$, which had $I^b_{b_{0}} = 15.19$ mV.

An extensive investigation of the critical neurons subjected 
to stimulations with currents in the range $I^\mathrm{stim} \in [14.5:16.0]$ mV
revealed that PBs can be observed only if the neurons $b_1$ and $b_2$ had
excitabilities within a narrow range (of amplitude $\simeq 0.2$ mV) centered around 
the threshold value. While, the stimulation of neuron 
$b_0$ revealed an absolute minimum in the PB activity
(an {\it anti-resonance}) at $I^\mathrm{stim} = 15.32$ mV and a relative minimum at 
$I^\mathrm{stim} = 15.45$ mV (see Fig.~S2 D). We performed also an 
extensive analysis of the response under SNS experiments for all the neurons of the network. 
As it is summarized in Fig.~S2 E, we found only sporadic significant 
reduction in the number of PBs occurring in extremely narrow current intervals. 

Therefore also for $N=200$ we observed that SND or SNS were capable to silence
the network and that this occurred only for a quite limited number of neurons,
which had low $K^T$ and reasonably high $I^b$. 
 
The detailed investigation of the burst events revealed that each PB was
always preceeded by the firing of the 3 critical neurons in the following order: 
$b_0 \to b_1 \to b_2  \to$ PB, as it is shown in Fig.~S2 FG.
The neuron $b_0$, which was the only one supra-threshold, fired first 
followed by the others (sub-threshold) and this triggered the onset of the PB.
As shown in Fig.~S2 D, whenever $b_1$ and $b_{0}$ were stimulated with currents 
$I^{stim} > I^b_{b_{0}} = 15.19$ mV the bursting activity stopped. 
This behaviour is analogous to what reported for the smaller network, 
indicating that neuron $b_{0}$ is the leader and the other ones are 
simply followers in the
construction of the PB, they cannot fire more rapidly then neuron 
$b_{0}$ or the PBs ceased. An analysis of the structural connectivity 
revealed that neuron $b_0$ projected an efferent synapse on $b_1$, 
which projected on $b_2$. We can safely affirm, also in this case, 
that neurons $b_0$, $b_1$ and $b_2$ form a functional clique, 
whose sequential activation is essential for the population burst onset.

\end{document}